\documentclass[notitlepage,twocolumn,prx,nobalancelastpage,amssymb,showpacs, superscriptaddress]{revtex4-2}

\usepackage{amssymb}
\usepackage{amsmath}
\usepackage{color}
\usepackage{graphicx}
\usepackage{xcolor}
\usepackage{hyperref}
\usepackage{float}

\newcommand{\be}{\begin{equation}}
\newcommand{\ee}{\end{equation}}
\newcommand{\bea}{\begin{eqnarray}}
\newcommand{\eea}{\end{eqnarray}}

\newcommand{\Ket}[1]{|#1\rangle}

\renewcommand{\epsilon}{\varepsilon}

\newcommand{\addMR}[1]{\textcolor{red}{#1}}

\newcommand{\addGK}[1]{\textcolor{green}{#1}}

\graphicspath{{figures/}}

\begin{document}

\title{
Vari-Cool: a non-unitary quantum variational protocol for simulated cooling
}
\author{Jeffrey Z. Song}
\thanks{These authors contributed equally to this work.}
\affiliation{Department of Physics, University of Washington, Seattle, WA 98195-1560, USA}
\author{Gilad Kishony}
\thanks{These authors contributed equally to this work.}
\affiliation{Department of Condensed Matter Physics,
Weizmann Institute of Science,
Rehovot 76100, Israel}
\author{Erez Berg}
\affiliation{Department of Condensed Matter Physics,
Weizmann Institute of Science,
Rehovot 76100, Israel}
\author{Mark S.~Rudner}
\affiliation{Department of Physics, University of Washington, Seattle, WA 98195-1560, USA}

\begin{abstract}
We introduce a variational approach for preparing low energy states of arbitrary target Hamiltonians. 
The protocol is defined in terms of a repeated cycle consisting of $p$ layers of unitary gates applied to the system and ancilla ``bath'' qubits, followed by reset of the bath qubits.
The gate parameters within each cycle are optimized such that the steady state achieved after many cycles has a low energy expectation value with respect to the target Hamiltonian, and that the energy converges toward the steady state value in as few cycles as possible.
We illustrate the protocol for the transverse field Ising model, and study its systematic behaviors with respect to system size, model parameters, and noise using tensor network based classical simulations.
We then experimentally demonstrate its operation on IBM's {\tt ibm\_kingston} quantum processor for up to 28 system qubits coupled to 14 bath sites.
Classical training on small system sizes and with few unitary layers per cycle gives robust results that transfer well to larger system sizes and to noisy hardware.
\end{abstract}

\maketitle

\section{Introduction}

A variety of tasks in quantum simulation~\cite{2012CiracZoller, Georgescu_2014}  and quantum computation~\cite{ZhouQAOAReview, Moll_2018, LanyonQCReview, BauerQC} require the preparation of ground or low-energy states of a given Hamiltonian.
Broadly speaking, the operating principle behind traditional approaches to ground state preparation such as adiabatic quantum computation~\cite{farhi2000quantum, Childs_2001} (i.e., ``quantum annealing''), the quantum approximate optimization algorithm (QAOA)~\cite{farhiQAOA, ZhouQAOAReview}, and the variational quantum eigensolver~\cite{Tilly_2022} is to design a unitary evolution that maps a simple, particular initial state on a set of qubits to a final state in which the desired ground state is approximately realized (possibly on a subset of the qubits, if ancilla qubits are included).

Recently, many groups have begun to explore the new opportunities and additional power afforded by incorporating {\it non-unitary} operations such as mid-circuit qubit measurements and resets into state preparation circuits~\cite{Park2016, Alhambra2019, Kaplan2017, Mazzola2019, Ferguson2021, Benfenati2021, mathhies2022adibatic_demag, Cubitt2023, Piroli2024, Cobos2024, Molpeceres2025, kishony2025gauged, kishony2025_chiral, mi2023stable, Lloyd2025, Ding2024, DingEndtoEnd, Zhan2025, Marti2025, langbehn2025}. 
In particular, physics-inspired cooling approaches use ancilla qubits with repeated resets to mimic the behavior of a low-temperature bath that can extract entropy and energy from the ``system qubits,'' analogous to natural cooling and thermalization processes.
With non-unitary circuits, protocols may be designed to prepare target states after a pre-determined fixed evolution, or as unique steady states of repeatedly applied cycles~\cite{mathhies2022adibatic_demag, mi2023stable, kishony2025gauged, Lloyd2025, Molpeceres2025}.



In this work we introduce a variational cooling (``Vari-Cool'') protocol for realizing low-energy states of arbitrary target Hamiltonians as the steady states of low-depth non-unitary cycles amenable to implementation on existing and near-term noisy intermediate scale quantum (NISQ) devices.
For illustration we focus on preparing approximate ground states of the transverse field Ising model (TFIM) in one dimension, which has served as a popular benchmark used in previous studies~\cite{Wiersema2020, mathhies2022adibatic_demag, kishony2025gauged, mi2023stable}.
We show how simulated cooling protocols such as that studied in Ref.~\cite{mathhies2022adibatic_demag}, 
which used on the order of 100 Suzuki-Trotter steps per cycle, can be compressed down to an optimized circuit with depth equivalent to three first-order Trotter steps, with only modest increases of the steady state energy (whereas naive truncation of continuous evolution down to just a few ``large'' Trotter steps does not provide a steady state with low energy relative to the ground state). 

We train the protocol classically on a tractable system of $N = 4$ system qubits coupled to $n_{\rm bath} = 2$ ancilla ``bath'' qubits via exact evolution, and observe good transferability to larger system sizes up to $N = 28$ system qubits coupled to $n_{\rm bath} = 14$ bath qubits using matrix product state (MPS) based simulations.
Through these simulations we explore the performance of the protocol across the TFIM phase diagram, and investigate the impact of gate errors.
We then experimentally demonstrate the protocol on IBM's {\tt ibm\_kingston} quantum processor~\cite{IBM_Quantum} on up to $N = 28$ system qubits connected to $n_{\rm bath} = 14$ bath sites, 
achieving performance comparable to that recently demonstrated by the Google team~\cite{mi2023stable}. 

The remainder of the paper is organized as follows.
In Sec.~\ref{sec:protocol} we describe the Vari-Cool protocol.
Then in Sec.~\ref{sec:training} we discuss training strategies.
In Sec.~\ref{sec:MPS} we present our classical simulations and results for ground state preparation of the TFIM.
In Sec.~\ref{sec:QPU} we present the experimental demonstration of the protocol on IBM's {\tt ibm\_kingston} quantum processor.
Finally, in Sec.~\ref{sec:Discussion} we summarize our results and discuss future directions.

\begin{figure*}[t!]
    \centering
    \includegraphics[width=\textwidth]{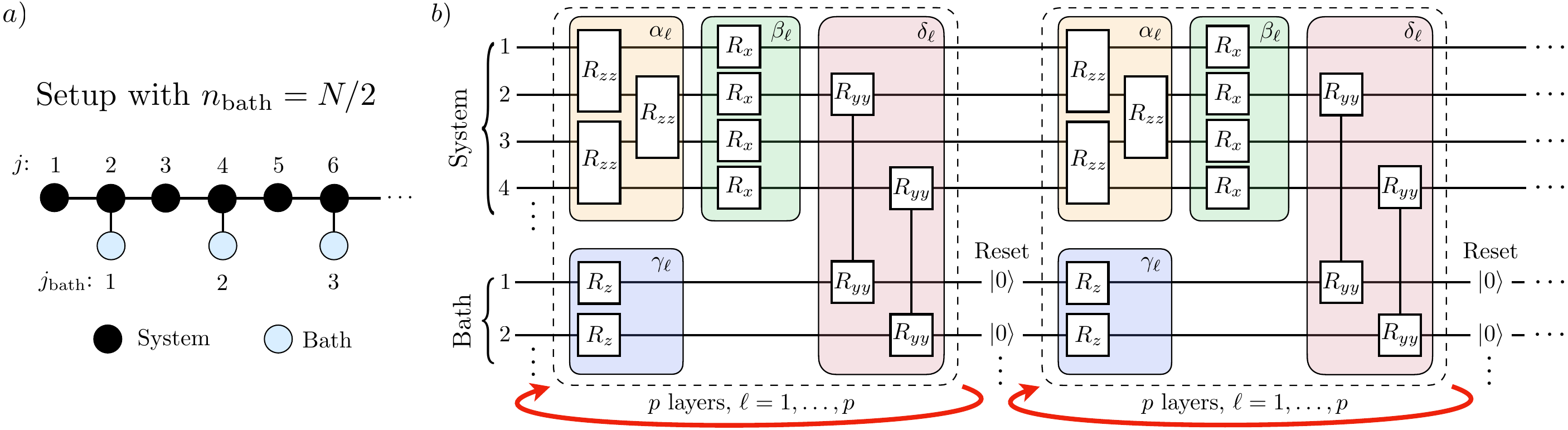}
    \caption{The Vari-Cool state preparation protocol, as applied to the transverse field Ising model.
    a) Setup and labeling for the case of $N$ system qubits coupled to $n_{\rm bath} = N/2$ bath qubits.
    The choice $n_{\rm bath} = N/2$ is convenient for the experimental setup in Sec.~\ref{sec:QPU}, but is not fundamental to the protocol.
    b) Low energy states are prepared by repeatedly applying a non-unitary cycle composed of unitary gates acting on system and ancilla ``bath'' qubits, and bath qubit reset operations.
    Within each cycle, $p$ layers of unitary gates are applied as shown, with rotation angles $\alpha_\ell$, $\beta_\ell$, $\gamma_\ell$, and $\delta_\ell$ chosen to produce a steady state (achieved after many cycles) with low energy evaluated with respect to the system Hamiltonian $\hat{H}_{\rm sys}$.
    }
    \label{fig:setup}
\end{figure*}
\section{The variational cooling protocol}\label{sec:protocol}

The Vari-Cool protocol is a flexible approach for approximate ground state preparation, which can be applied to a wide variety of target Hamiltonians.
The setup consists of $N$ system qubits connected to $n_{\rm bath}$ ``bath'' sites,
see Fig.~\ref{fig:setup}a.
For efficient cooling and greatest robustness to gate errors, we consider $n_{\rm bath}$ to be an $\mathcal{O}(1)$ fraction of $N$, but in principle it need not be.
Throughout this work we will assume a fixed layout of the qubits, utilizing only local couplings between system and bath qubits; nonlocal couplings could also be considered, particularly for implementations on trapped ion or neutral atom platforms where arbitrary connections are possible.


As shown in Fig.~\ref{fig:setup}b, the protocol runs as a cycle composed of a block of unitary gates that act on the system and bath qubits, followed by resets of the bath qubits.
For concreteness, we assume the bath qubits to always be reset to the $\Ket{0}$ state.
Repeating this cycle drives the system qubits towards a steady state; we design and variationally optimize the cycle unitary block to give a low energy expectation value with respect to the target Hamiltonian in the steady state, and rapid convergence to the steady state with the number of cycles applied.

In principle, any parameterized unitary could be used for the unitary block.
Here we take inspiration from recently developed simulated cooling protocols~\cite{Park2016, Kaplan2017, Mazzola2019, mathhies2022adibatic_demag, Piroli2024, Molpeceres2025, kishony2025gauged, kishony2025_chiral, mi2023stable, Lloyd2025, Ding2024, DingEndtoEnd, Zhan2025, Marti2025, Shin2025}, which are based on a Trotterization of continuous time evolution of system and bath qubits: the system qubits evolve according the target Hamiltonian, $\hat{H}_{\rm sys}$, each bath qubit evolves according to a local on-site Hamiltonian, $\hat{H}_{\rm bath}$, and a system-bath coupling $\hat{H}_{\rm int}$ is used to transfer energy from the system to the bath. 
We construct the unitary block 
using $p$ unitary layers that mimic Trotter steps of such joint evolution. 
To enhance feasibility for implementation on NISQ devices, we assign independent variational parameters for gates corresponding to different terms in the system, bath, and system-bath interaction Hamiltonians, and for each layer $\ell = 1, \ldots, p$, and allow these unitaries to deviate substantially from the identity~\footnote{Note that the experiment of Ref.~\cite{mi2023stable} employed variational optimization with respect to the values of a constant (layer-independent) applied bath field and a constant system-bath coupling strength. The Vari-Cool protocol is a general purpose variational scheme in which the unitary block may or may not resemble Trotterized evolution, and where all parameters may vary independently on every layer.}.
The relationship between the Vari-Cool protocol with this ansatz and simulated cooling is analogous to that between the quantum approximate optimization algorithm (QAOA) and adiabatic quantum annealing, where the large number of Trotter steps needed for digital quantum annealing are replaced by a small number of variationally-parameterized layers of similar form to the original Trotter steps.


For a large number of layers, $p$, the Vari-Cool protocol is expressive enough to perform at least as well as programmable adiabatic demagnetization \cite{mathhies2022adibatic_demag} or other simulated cooling protocols; if trained to run for a single cycle, it can also capture the behavior of standard VQE by setting the system bath coupling parameters to zero.
When the number of layers is small, the unitary blocks may no longer resemble Trotterized evolution under any smoothly changing Hamiltonian, but nonetheless we anticipate (and will confirm below) that it can still prepare good low-energy states with respect to the target Hamiltonian, $\hat{H}_{\rm sys}$.
In the subsection below we will discuss the specific setup and application of the protocol to the transverse field Ising model, then discuss its training in Sec.~\ref{sec:training}.

\subsection{Application to the transverse field Ising model}

As a concrete demonstration, we apply our variational cooling protocol to the transverse field Ising model (TFIM) on $N$ spins in a one dimensional chain.
The system qubits represent the spins of the TFIM, with system Hamiltonian
\be
\label{eq:TFIM} \hat{H}_{\rm sys} = -J\sum_{j=1}^{N-1} \hat{Z}_j\hat{Z}_{j+1} - h \sum_{j = 1}^N \hat{X}_j,
\ee
where $\hat{X}_j,\hat{Y}_j,\hat{Z}_j$ are the Pauli operators acting on spin (qubit) $j$, $J > 0$ is the (ferromagnetic) Ising exchange coupling, and $h$ is the transverse field strength.
In Eq.~(\ref{eq:TFIM}) and throughout this work we take open boundary conditions for the chain.

In anticipation of the experimental implementation (see Sec.~\ref{sec:QPU} below), we tailor the circuit to fit the 
qubit connectivity offered by the {\tt ibm\_kingston} processor~\footnote{Ideally, $n_{\rm bath} = N$ would be more effective for cooling the system qubits.
However, the additional overhead that would be needed to introduce bath couplings to every system site with the restricted connectivity shown in Fig.~\ref{fig:QPU} would far outweigh the benefits.}. 
In particular, as shown in Fig.~\ref{fig:setup}a, we choose $n_{\rm bath} = N/2$ ($N$ will always be taken to be even), with system qubits on even sites $j = 2, 4, \ldots, N$ coupled to bath sites $j_{\rm bath} = 1, 2, \ldots, n_{\rm bath}$.

Each unitary block of the protocol consists of $p$ layers composed of single and two-qubit gates as shown in Fig.~\ref{fig:setup}b.
The unitary transformation for a given layer $\ell$ is given by 
\be
\label{eq:block_unitary} \hat{U}_\ell = \hat{R}_{yy}(\delta_\ell)\, \hat{R}_{z}(\gamma_\ell)\, \hat{R}_{x}(\beta_\ell)\,  \, \hat{R}_{zz}(\alpha_\ell),
\ee
with 
\bea
\nonumber \hat{R}_{zz}(\alpha_\ell) &=& e^{-i (\alpha_\ell/2) \sum_{j = 1}^{N-1} \hat{Z}_j\hat{Z}_{j+1} }\\
\nonumber \hat{R}_x(\beta_\ell) &=& e^{-i (\beta_\ell/2) \sum_{j=1}^N \hat{X}_j}\\
\nonumber \hat{R}_{z}(\gamma_\ell) &=& e^{-i (\gamma_\ell/2) \sum_{j = 1}^{N/2} \hat{Z}_{{\rm bath},j}}\\
\label{eq:block_unitary_decomposed}\hat{R}_{yy}(\delta_\ell) &=& e^{-i (\delta_\ell/2) \sum_{j = 1}^{N/2}\hat{Y}_{{\rm bath}, j}\hat{Y}_{2j}}
\eea
Here $\hat{X}_{{\rm bath}, j},\hat{Y}_{{\rm bath}, j},\hat{Z}_{{\rm bath}, j}$ are the Pauli operators on bath qubit $j$.
To keep the number of variational parameters low, we use common rotation angles for all site/bonds within a given operation.
This translation-invariant ansatz also promotes transferability from smaller to larger system sizes. 

After each unitary block of $p$ layers is applied, the bath qubits are reset to the $\Ket{0}$ state.
The full cycle of unitary evolution followed by bath reset defines a quantum channel $\mathcal{E}$ that acts on the system qubits.
After many cycles of the protocol, the system tends to a steady state $\hat{\rho}_{\rm steady}$ satisfying $\mathcal{E}(\hat{\rho}_{\rm steady}) = \hat{\rho}_{\rm steady}$.
The goal is then to optimize the parameters $\boldsymbol{\alpha}, \boldsymbol{\beta}, \boldsymbol{\gamma}, \boldsymbol{\delta}$ 
to minimize the expectation value of the system's energy in the steady state, $E_{\rm steady}(\boldsymbol{\alpha}, \boldsymbol{\beta}, \boldsymbol{\gamma}, \boldsymbol{\delta}) = {\rm Tr}[\hat{\rho}_{\rm steady} \hat{H}_{\rm sys}]$;
here $\boldsymbol{\alpha} = \{\alpha_1, \ldots, \alpha_p\}$, and similarly for $\boldsymbol{\beta}, \boldsymbol{\gamma}$, and $\boldsymbol{\delta}$.


\section{Training on classical hardware}\label{sec:training}
The goals of training are to optimize the circuit's parameters to:
\begin{enumerate}
 \item[(i)] achieve a low steady state energy, $E_{\rm steady}$, with a small number of layers, $p$, per cycle,
 \item[(ii)] achieve rapid cooling, such that energy converges toward $E_{\rm steady}$ in as few cycles as possible.
\end{enumerate}    
Keeping the numbers of required layers and cycles small is particularly important for implementation on NISQ hardware, where errors can build up for large depth circuits.
The non-unitary protocol exhibits an inherent robustness to noise through transfer and elimination of excitations through the bath qubits; the steady state is most resilient to hardware noise when cooling is rapid and the depth $p$ of each unitary block is small.


Variational algorithms are challenging to train on NISQ devices due to the presence of both device noise and measurement shot noise, which is costly due to the necessity of using large numbers of measurement shots per training round and poses difficulties for the classical routines needed to find optimal values of variational parameters~\cite{Cerezo2021}.
Here we focus on quantum simulation of phases of matter in translation-invariant systems.
In this context, we train the protocol classically on small-sized systems using exact evolution of the density matrix with no added noise, and demonstrate transferability to larger system sizes with and without noise via classical MPS based simulations (Sec.~\ref{sec:MPS}) and experimentally on the {\tt ibm\_kingston} device (Sec.~\ref{sec:QPU}).
The protocol could also be trained in a traditional hybrid mode (with energies estimated experimentally on a quantum device), or in a combination where classical training on a small system size is used to produce a good initial guess for further hybrid training on the quantum device.

We have explored several approaches for training our protocol.
In the main text we will outline the most important considerations and most relevant aspects for the regime of interest in this study.
In Appendix~\ref{app:training} we will give further details and discuss alternative approaches that we also found effective.

A key difference between training the Vari-Cool protocol and other variational algorithms such as the variational quantum eigensolver (VQE) is that here we wish to optimize for the {\it steady state} achieved after many cycles, rather than the state output from a circuit of fixed length, and we aim to further ensure rapid convergence to the steady state [objective (ii) above].
Training on the steady state itself (approximated, for example, by the state of the system after several tens of cycles) is a natural and effective choice for obtaining a low steady state energy.
However, that approach by itself does not enforce rapid cooling and can lead to long cooling times.
Another natural choice is to train for the minimal energy after one cycle (either for a fixed initial state or averaged over some ensemble of initial states).
We have observed that this approach often yields optimized parameters that give poor steady state energies, or can lead to unsteady or limit cycle type behavior where the state of the system does not converge to a steady state after many cycles (see Appendix~\ref{app:training}).

Considering the above, we found that training based on the expectation value of the system's energy after several cycles achieves the dual aims of obtaining low steady state energies and rapid cooling.
We will refer to the number of cycles after which the energy is evaluated for training as $T_{\rm train}$.
To avoid unsteady long time behavior, we further impose a constraint that at each training iteration the optimizer only accepts sets of parameter values such that the energy monotonically decreases over a window of $2\tau + 1$ cycles surrounding $T_{\rm train}$ (see Appendix~\ref{app:training} for further details).

\begin{figure}[t]
    \centering
    \includegraphics[width=\linewidth]{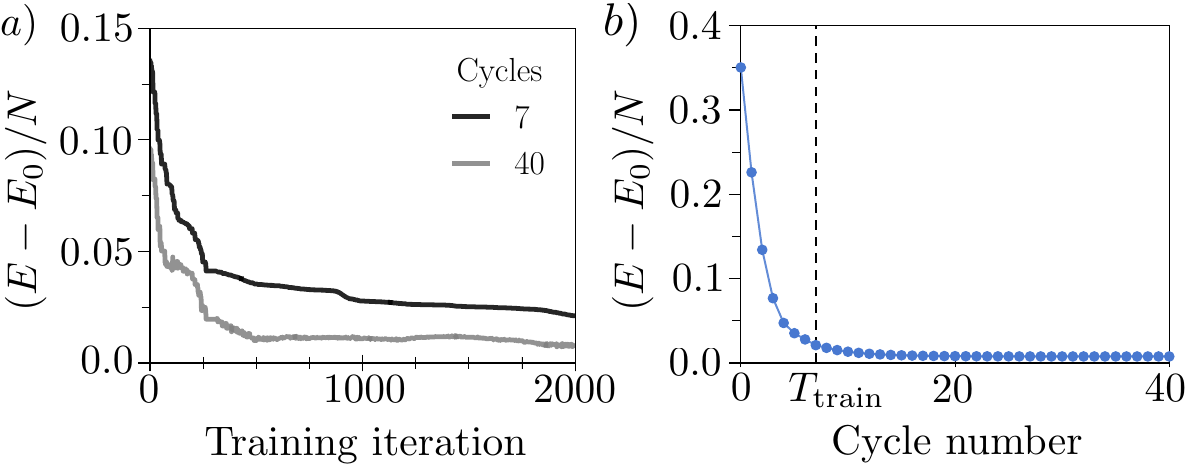}
    \caption{
    Example of training the cooling circuit.
    Parameters are optimized on a classical noiseless simulator with $N = 4$ system qubits and $n_{\rm bath} = 2$ bath qubits, and $p = 3$ layers in the unitary block.
    Here we show the case $J = 0.4$, $h = 0.6$.
    a) 
    The black solid line shows the %
    residual energy density $(E - E_0)/N$ after $T_{\rm train} = 7$ cycles; this value is used for the optimization in each training iteration. Here $E_0 = -2.6016$ is the ground state energy (in the units above where $J + h = 1$).
    The gray line (below the black line) shows the steady state energy density achieved by applying the circuit 40 times with the current values of the parameters at each training iteration.
    See Sec.~\ref{sec:training} for details of the training.
    b) Residual energy density 
    as a function of the number of cycles applied, obtained with the final optimized parameters from the training in panel a.
    The system qubits were initialized in the state $\Ket{0000}$.
    The residual energy density after $40$ cycles is 
    $(E - E_0)/N = 7.7 \times 10^{-3}$ per site.
    } 
    \label{fig:Training}
\end{figure}
For the demonstrations below, we trained the protocol for the transverse field Ising model with the translation-invariant ansatz in Fig.~\ref{fig:setup}b and Eqs.~(\ref{eq:block_unitary}) and (\ref{eq:block_unitary_decomposed}) on a system with $N = 4$ system qubits with open boundary conditions and $n_{\rm bath} = 2$ bath qubits, 
using $T_{\rm train} = 7$ cycles and $\tau = 2$ cycles.
We used the Nelder-Mead method~\footnote{The Nelder-Mead method is a gradient-free approach for minimizing or maximizing multidimensional cost functions. See \href{https://docs.scipy.org/doc/scipy/reference/optimize.minimize-neldermead.html}{SciPy documentation} and reference therein for further information.} 
to find the optimal parameters.
In Fig.~\ref{fig:Training}a we show an example of how the energy after $T_{\rm train} = 7$ cycles evolves as training progresses; we also show the value of the energy after 40 cycles, which serves as a proxy for the steady state energy achieved using the current values of the parameters at each iteration of the training.
In Fig.~\ref{fig:Training}b we show how energy decreases toward the steady state value  as a function of the number of cycles applied with the optimized parameters found in Fig.~\ref{fig:Training}a. 

Each layer of the protocol's unitary block contributes four variational parameters $\alpha_\ell, \beta_\ell, \gamma_\ell$, and $\delta_\ell$, as shown in Fig.~\ref{fig:setup}b.
Therefore, an implementation with $p$ layers contains $4p$ variational parameters.
We found that different approaches could be most effective, depending on the desired depth $p_*$. 
For relatively large depth unitary blocks with $p_* \gtrsim 10$,  a coarse Trotterization of a simulated cooling protocol such as the adiabatic demagnetization protocol in Ref.~\onlinecite{mathhies2022adibatic_demag} provides a good initial guess for the parameters $\{\alpha_\ell\}, \{\beta_\ell\}, \{\gamma_\ell\}$, and $\{\delta_\ell\}$.
For small values of $p_* \approx 3-6$, as desired here for implementability on a present-day quantum device, the naive ``coarse Trotter'' initialization of circuit parameters led to poor training results without further manual intervention. 

We now briefly outline 
two effective approaches that we developed for training the protocol to operate with a small number of layers, $p_*$; see Appendix~\ref{app:training} for further considerations.
One approach is to first train with large $p \approx 10$, where coarse Trotterization gives a good starting point, then successively prune off layers and re-train down to the desired final circuit size $p_*$.
Alternatively, for small enough $p_*$ it is sufficient to try a large number of random initializations of the $4p_*$ parameters, use the (approximate) steady state energy obtained after a fixed, large number of cycles (before any optimization) to identify and down-select to a handful of the most promising initial parameter sets~\footnote{Here it is important to also enforce a monotonic decrease of energy with cycle number around $T_{\rm train}$ to avoid unsteady behavior, as we impose throughout the optimization.}, then optimize from each of these initializations and choose the best.

The random initialization approach has the advantage that it is trivially parallelizable.
For $p_* = 3$ and 15,000 random initializations we consistently find initializations leading to good, converged results; Fig.~\ref{fig:Training} shows the training starting from one of the promising initializations identified from such a set of 15,000.
The pruning method offers a straightforward and systematic way to reduce the layer count while maintaining good steady state performance (see Appendix~\ref{app:training}). 
For the simulations and experiments below we use these training procedures to obtain optimized parameters at system size $N = 4$, $n_{\rm bath} = 2$ for $J = 0.4, h = 0.6$, $J = 0.45, h = 0.55$, $J = 0.55, h = 0.45$, and $J = 0.6, h = 0.4$.
The optimized parameter sets are given in {Tables I-IV} of Appendix~\ref{app:training}.



\section{Classical simulations}\label{sec:MPS}

We demonstrate the transferability of the translation-invariant cooling circuit trained on a small system size ($N=4$) to larger system sizes and assess the performance of the protocol in the presence of noise through stochastic MPS simulations using Qiskit \cite{javadiabhari2024quantumcomputingqiskit}. In these simulations, the initial pure state $\Ket{0}^{\otimes (N+n_{\rm{bath}})}$ represented as an MPS evolves deterministically through the unitary gates of the cooling cycle. Both noise and resets (equivalent to computational basis measurement and a conditional bit-flip correction), which are non-unitary quantum channels, are performed stochastically such that the state remains pure within each trajectory.

Within these simulations, we model hardware noise by inserting a two-qubit depolarizing channel parametrized by probability $\xi$ after every two-qubit gate (a uniformly random two-qubit Pauli gate is applied with probability $\xi$), and a single-qubit depolarizing channel parametrized by the probability $\xi/10$ after every single-qubit gate 
(a uniformly random one-qubit Pauli gate is applied with probability $\xi/10$).
A realistic value of $\xi$ for the hardware used in our experiments in Sec.~\ref{sec:QPU} is on the order of $10^{-2}$, which is at the upper end of the values used for our simulations in this section.

We simulate the cooling of the TFIM using the optimized circuits in both the paramagnetic phase ($J=0.4,h=0.6$ and $J=0.45,h=0.55$) and the ferromagnetic phase ($J=0.6,h=0.4$ and $J=0.55,h=0.45$) in the presence of varying levels of noise $\xi$. 
We study system sizes $N=4,8,16,28$, using $7000$, $3500$, $175$, and $100$ shots or trajectories of the simulation, respectively.
The MPS bond dimension is truncated to $64$ throughout all simulations. After convergence to the steady state (estimated as the state after $40$ cycles) we compute the energy with respect to $\hat{H}_{\rm{sys}}$ and spin-spin correlations of the form $\langle\hat{Z}_i\hat{Z}_j\rangle$.

\begin{figure}[t]
    \centering
    \includegraphics[width=\linewidth]{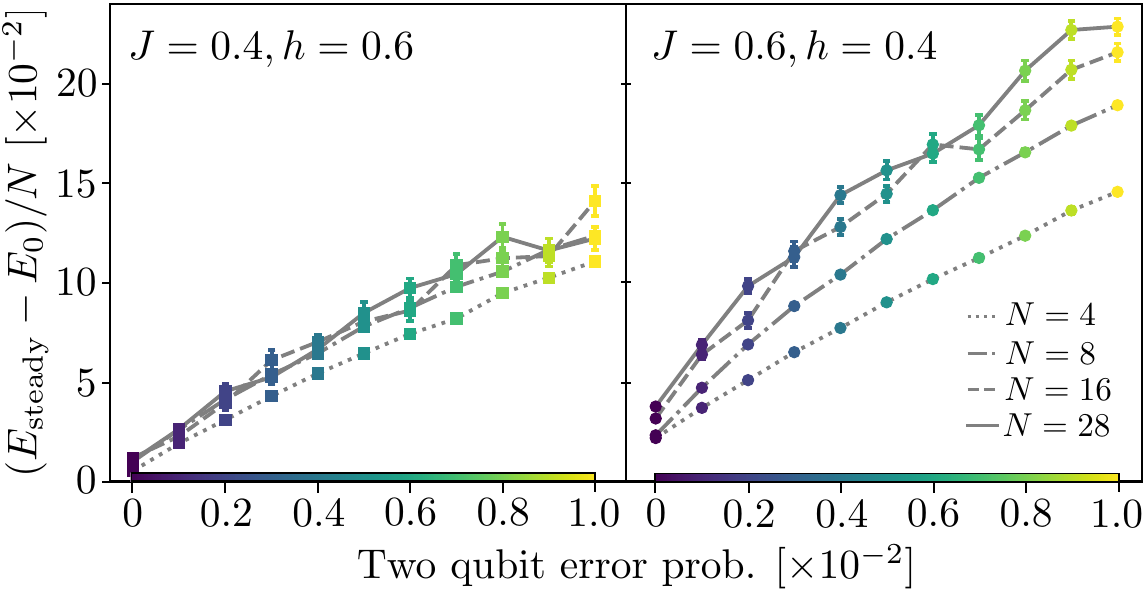}
    \caption{Dependence of the steady state energy density on two qubit gate error probability, from stochastic evolution of matrix product states. Different system sizes $N=4,8,16,28$ (with $n_{\rm{bath}}=N/2$) are represented by different line styles.
    Left panel: Energy density relative to the ground state, $(E_{\rm steady}-E_0)/N$, in the paramagnetic phase with $J = 0.4$, $h = 0.6$.
    Right panel: Energy density relative to the ground state in the ferromagnetic phase with $J = 0.6$, $h = 0.4$.
    The larger overall scale and system size dependence of the energy density in the ferromagnetic phase reflect the challenge of cooling topological domain wall excitations in the ferromagnetic phase~\cite{mathhies2022adibatic_demag}.
    }
    \label{fig:MPS_noise}
\end{figure}
In Fig.~\ref{fig:MPS_noise} we show the steady state energy density relative to the ground state of $\hat{H}_{\rm{sys}}$ 
as a function of noise $\xi$ for different system sizes, both in the paramagnetic phase ($J=0.4,h=0.6$) and in the ferromagnetic phase ($J=0.6,h=0.4$).
The error bars 
represent the square-root-sum-of-squares combination of the standard error due to the variance between shots of the stochastic simulation and the error due to finite bond dimension truncation. 
We estimate the truncation error as the absolute difference between an expectation value computed at bond dimension $64$ and the same value at bond dimension $32$.

In the limit of large system sizes, the energy density becomes system size independent in both phases. In this regime, the steady state represents a balance between a bulk cooling rate and a rate of heating due to noise \cite{mathhies2022adibatic_demag}. In the smaller system sizes, a lower steady state energy density is reached due to a combination of two effects. The first is the transfer of the parameters trained on system size $N=4$ directly to larger system sizes where they may not be optimal without further fine-tuning. The second is an intrinsic finite size effect of cooling at the edges of the system, which is more dominant at smaller system sizes. This system size dependence is more pronounced in the ferromagnetic case; the ferromagnetic phase is inherently more challenging to cool than the paramagnetic phase due to the topological nature of its domain wall excitations, which cannot be removed individually by a local process \cite{mathhies2022adibatic_demag}. In fact, at very low energy densities (below those reachable with the low-depth cycles used here) and in the thermodynamic limit, the energy density is expected to have a square-root noise dependence in the ferromagnet, and a linear dependence in a paramagnet. This is also the reason for the difference in overall scale of the energy densities reached for dual points in the two phases.

\begin{figure}[t]
    \centering
    \includegraphics[width=\linewidth]{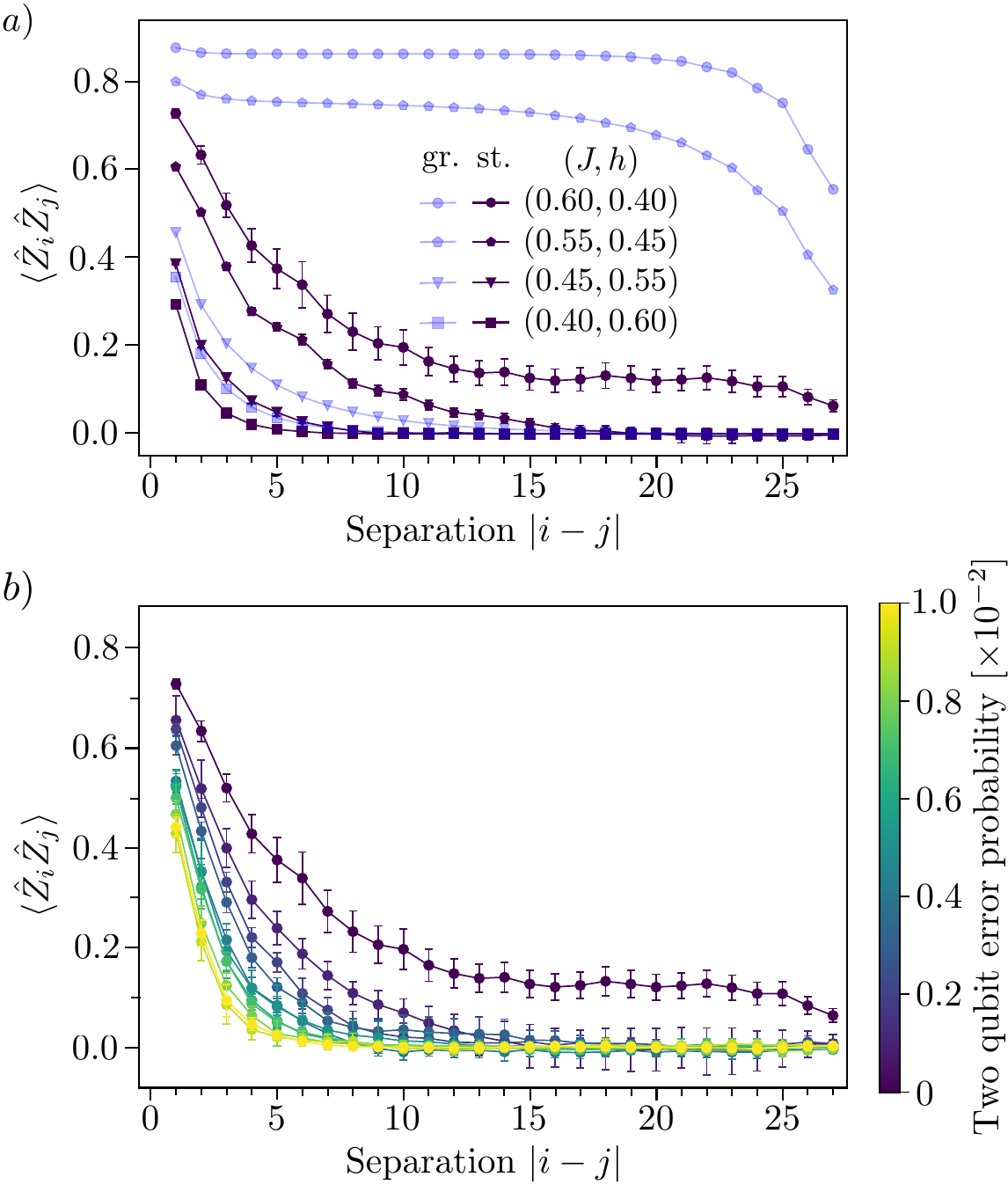}
    \caption{Steady state spin-spin correlations, from classical simulations.
    a) Spin-spin correlation function, at four representative points in the phase diagram at the steady state reached in the absence of noise (``st.,'' dark lines) and in the ground state (``gr.,'' light lines).
    The system reaches a broken symmetry state with finite magnetization in the ferromagnetic phase, resulting in the plateau seen for $J = 0.6, h = 0.4$.
    b) Spin-correlation function for various noise levels at $J = 0.6, h = 0.4$. The long-ranged ferromagnetic correlations are rapidly washed out by the presence of noise.
    }
    \label{fig:MPS_correlations}
\end{figure}
In Fig.~\ref{fig:MPS_correlations} we examine the spin-spin correlations, $\langle \hat{Z}_i\hat{Z}_j\rangle$ in the steady state vs.~the separation $|i-j|$ with $i$ and $j$ chosen symmetrically around the center of the TFIM chain ($\lfloor(i+j)/2\rfloor= N$)  for $N=28$. 
As above, the error bars capture variance of the mean due to both shot noise of the stochastic simulations and due to bond dimension truncation. Figure~\ref{fig:MPS_correlations}a shows these correlations for the different points in the $(J,h)$ phase diagram in the absence of gate noise ($\xi=0$); we show the spin-spin correlations in the ground state for reference (light color).
As expected, in the ferromagnetic phase, especially at $J=0.6,h=0.4$ (further away from the critical point), the correlations are long-ranged, while in the paramagnet they decay rapidly.
Continuing with the parameters $J=0.6,h=0.4$, in Fig.~\ref{fig:MPS_correlations}b we show the correlations with varying levels of noise. The long range correlations are degraded by the noise due to the proliferation of domain wall excitations.
A high sensitivity of the long-range correlations to noise is expected due to the fact that domain walls are topological excitations that can only be removed from the bulk of the system in pairs~\cite{mathhies2022adibatic_demag}.
Alternative forms for the unitary block of the protocol, for example inspired by the ``gauged cooling'' approach described in Ref.~\onlinecite{kishony2025gauged}, may enable more efficient cooling and stabilization of long-range correlations.



\section{Experimental demonstration}\label{sec:QPU}

We now discuss our experimental implementation and results using IBM's {\tt ibm\_kingston} Heron R2 quantum processor.
We illustrate the protocol's performance both at $J = 0.4, h = 0.6$ and $J = 0.45, h = 0.55$, using the optimized parameters found from classical training at $N = 4, n_{\rm bath} = 2$ as described above. 
In the main text we focus on results for $N = 6, n_{\rm bath} = 3$ and $N = 28, n_{\rm bath} = 14$ sites, with $p = 3$ layers in the unitary block of the protocol.
Conveniently, $\hat{R}_{zz}(\theta)$ for arbitrary $\theta$ [as used in our unitary block, see Eqs.~(\ref{eq:block_unitary}) and (\ref{eq:block_unitary_decomposed})] is a native ``fractional'' gate on the {\tt ibm\_kingston} processor; this helps to keep the gate count low~\footnote{Each $\hat{R}_{yy}(\theta)$ gate in the protocol is transpiled to a native $\hat{R}_{zz}(\theta)$ gate and single qubit gates.}. 
Details of the qubit mappings to the device are specified in Appendix~\ref{app:QPU}.
In Appendix \ref{app:QPU} we also discuss results for a second geometry with $N = 6$ (measured on the same chip, in parallel with the $N = 6$ results shown in the main text), as well as results for $p = 4$ layers per unitary block.

\begin{figure}[t]
    \centering
   \includegraphics[width=\columnwidth]{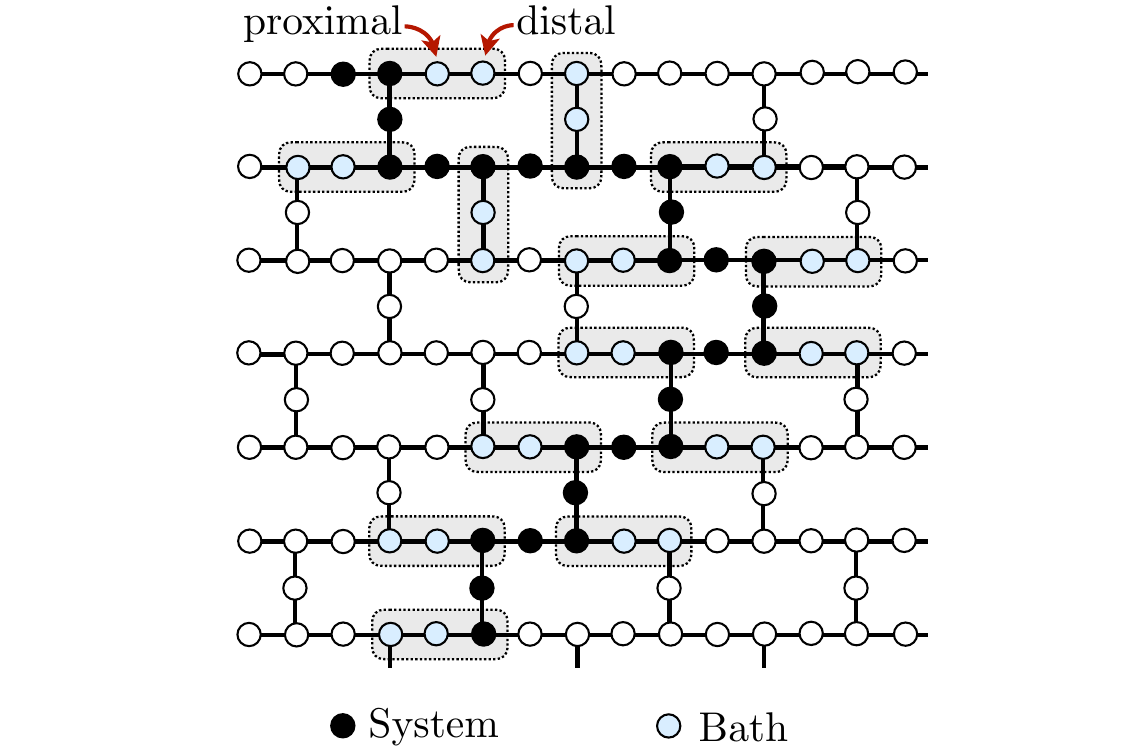}
    \caption{
    Qubit layout for experimental runs on the {\tt ibm\_kingston} quantum processor.
    To minimize waiting times and potential cross-talk errors during bath qubit {\tt RESET} operations on the quantum processor, we use two qubits for each bath site.
    At the {\tt RESET} step of the protocol in Fig.~\ref{fig:setup}b, we apply a {\tt SWAP} operation between the proximal and distal bath qubits (see labeling on figure), then apply a {\tt RESET} to the distal bath qubit in parallel with the next unitary block.
    }
    \label{fig:QPU}
\end{figure}
To mitigate errors incurred during slow qubit reset operations, we employ two qubits per bath site as shown for the $N = 28, n_{\rm bath} = 14$ layout in Fig.~\ref{fig:QPU}.
We refer to the bath qubit nearest to and directly coupled to the system chain as the ``proximal'' bath qubit, and the other as the ``distal'' bath qubit.

The unitary gates act on the system and proximal bath qubits, precisely as described above and depicted in Fig.~\ref{fig:setup}.
Initially the distal bath qubits are prepared in the $\Ket{0}$ state.
Rather than simply resetting the proximal bath qubits, which imposes a significant delay between cycles and may induce additional cross-talk errors to nearby system qubits~\footnote{We have not systematically studied the effects of cross-talk during resets, but did find overall that the two qubits per bath site approach described in Fig.~\ref{fig:QPU} gave better results than a naive direct implementation with one qubit per bath site.}, at the {\tt RESET} step of the protocol we apply a {\tt SWAP} operation between the distal and proximal bath qubits~\footnote{{\tt SWAP} gates are not native to the {\tt ibm\_kingston} processor. Instead, a {\tt SWAP} operation can be transpiled to a sequence of three {\tt CX} gates acting in alternating directions between the two qubits. In our protocol, however, one of the qubits involved in each {\tt SWAP} operation always begins in the $\Ket{0}$ state, rendering the first of the three {\tt CX} gates redundant. Thus only two {\tt CX} gates are required. Finally, each {\tt CX} gate can be implemented using one native {\tt CZ} gate and single qubit gates.}.
This {\tt SWAP} refreshes the state of the proximal bath qubit to $\Ket{0}$ as desired; we then apply a {\tt RESET} to the distal qubit in {\it parallel} with the next unitary block (which again acts only on the system and proximal bath qubits).
This avoids delays during the {\tt RESET} and keeps the dissipative operations further away from the system qubits.
Due to the ancillas used in each bath, in total we used 12 qubits for system size $N = 6$ and 56 qubits for system size $N = 28$.

\begin{figure}[t!]
    \centering
    \includegraphics[width=\columnwidth]{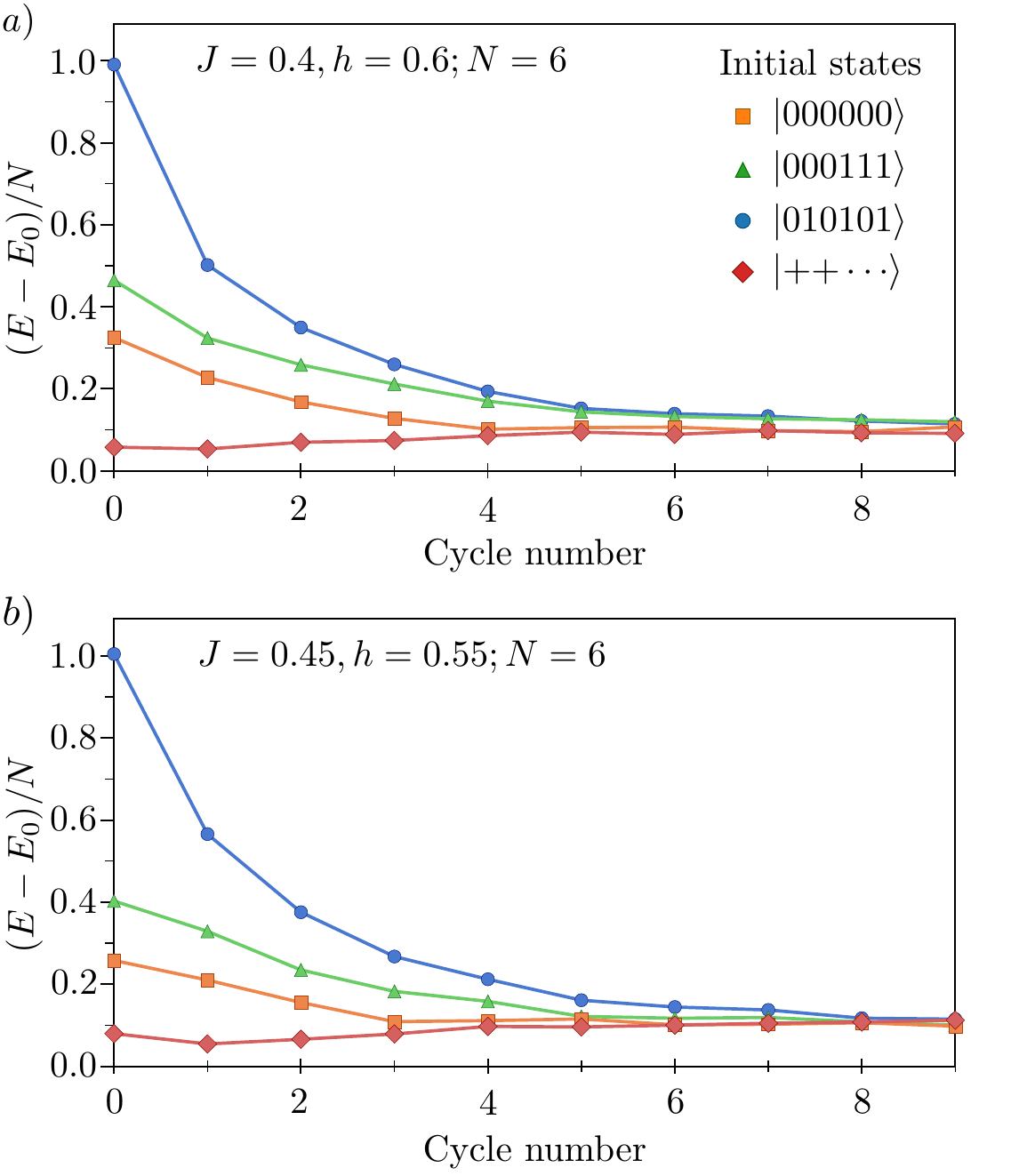}
    \caption{Experimental demonstration of the Vari-Cool protocol on the {\tt ibm\_kingston} quantum processor with $N = 6$ system sites and $n_{\rm bath} = 3$ two-qubit baths, and $p = 3$ layers per cycle.
    The average energy density relative to the ground state is shown as a function of the number of cooling cycles for several initial states as indicated on the legend.
    Averages are obtained from 8192 shots per measurement basis ($Z$ or $X$), for each fixed number of cycles.
    The protocol is run with the same parameters obtained from classical training on $N = 4$, $n_{\rm bath} = 2$ as used in the classical MPS-based simulations in Sec.~\ref{sec:MPS}.
    a) Results for $J=0.4, h = 0.6$, with ground state energy $E_0 = -3.9390$. 
    Based on the observed trajectories, we find a steady state residual energy above the ground state of $0.101 \pm 0.011$ per system site.  
    b) Results for $J = 0.45, h = 0.55$, with ground state energy $E_0 = -3.7720$. 
          Here we find a steady state residual energy above the ground state of $0.106\pm 0.007$ per system site.
    }
    \label{fig:Results_6+3}
\end{figure}
In Fig.~\ref{fig:Results_6+3} we show the expectation value of the system residual energy density (relative to the ground state) with respect to the TFIM target Hamiltonian for a system of $N = 6$ qubits, estimated from 8192 measurement shots per measurement basis ($Z$ or $X$), after 0 to 9 cycles for four different initial states as indicated on the legend.
Error bars estimated from the standard error cover ranges smaller than the plot markers.

In both panels, it is clear that the state of the system approaches a steady state (independent of the initial state) with a characteristic time scale of 2-3 cycles.
The residual energy is slightly higher for $J = 0.45, h = 0.55$, presumably due to the closer proximity of this parameter set to the critical point of the TFIM.
In both cases, the steady state captures $E/E_0 \gtrsim 80\%$ of the ground state energy.


\begin{figure}[t]
    \centering
    \includegraphics[width=\linewidth]{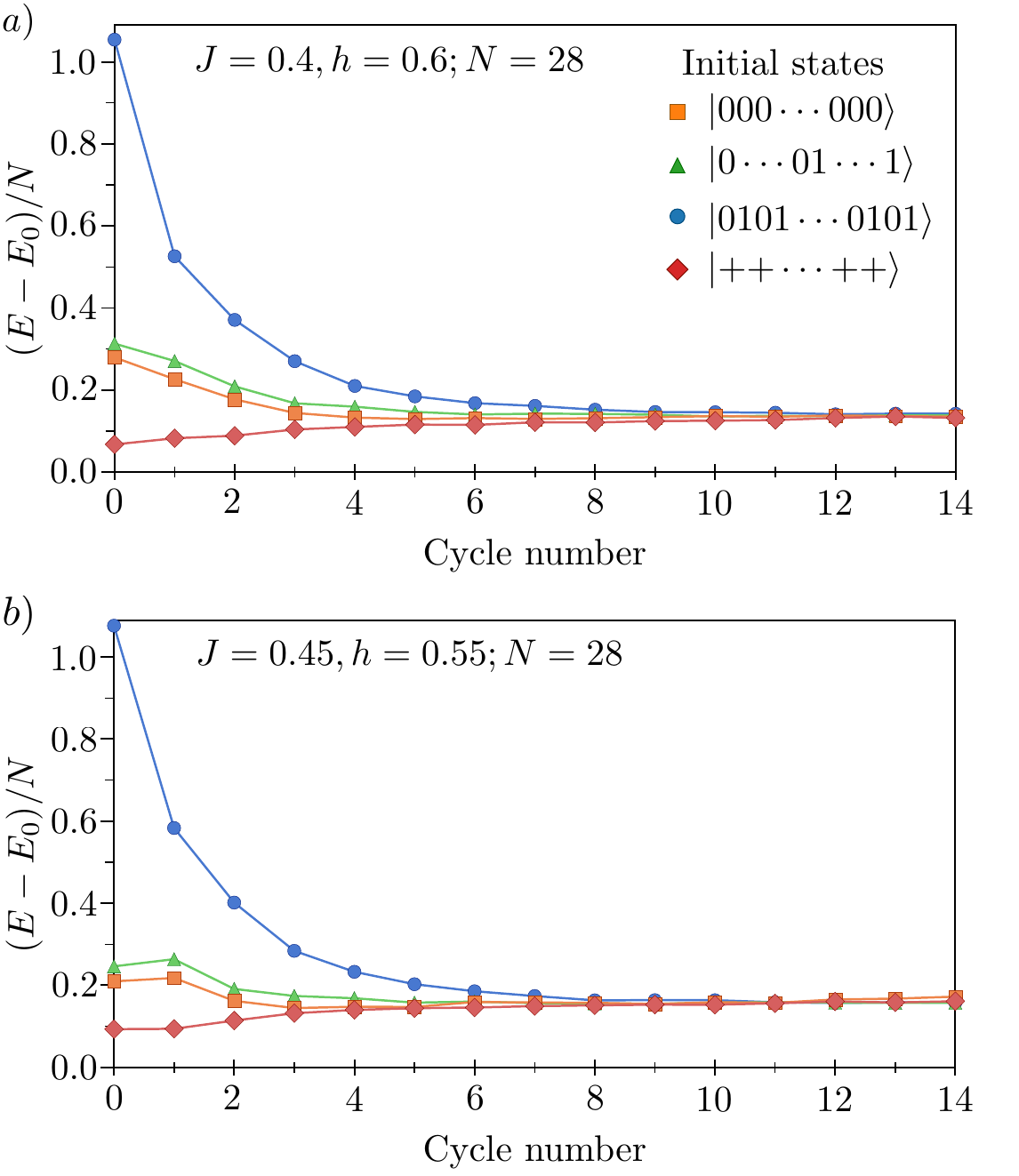}
    \caption{Experimental demonstration of the Vari-Cool protocol for $N = 28$ system qubits coupled to $n_{\rm bath} = 14$ two-qubit baths on the {\tt ibm\_kingston} quantum processor, with layout as shown in Fig.~\ref{fig:QPU}.
    The average energy density (relative to the ground state) as a function of the number of cooling cycles is shown for several initial states as indicated on the legend.
    Averages are obtained from 8192 shots per measurement basis ($Z$ or $X$), for each fixed number of cycles.
    a) Results for $J=0.4, h = 0.6$, with ground state energy $E_0 = -18.6520$.
    The observed trajectories give an average steady state residual energy density of $0.137 \pm 0.004$ per system site. 
    b) Results for $J=0.45, h = 0.55$, with $E_0 = -18.0014$.
    Here, the average residual energy in the steady state is $0.162 \pm 0.006$ per system site.
    }
    \label{fig:Results_28+14}
\end{figure}

Next we demonstrate the transferability of classical training on a small system to a large, noisy quantum device.
In Fig.~\ref{fig:Results_28+14} we show the expectation value of the system's residual energy density above the ground state 
for a system of $N = 28$ system qubits, estimated from 8192 measurement shots per measurement basis ($Z$ or $X$), after 0 to 14 cycles for four different initial states as indicated on the legend.
At $N = 28$ system qubits, and with simple training on a classical computer with a translation invariant ansatz with just $N = 4$ system qubits, the protocol recovers 
$E/E_0 \approx 79\%$ of the ground state energy for $J = 0.4, h = 0.6$, and $E/E_0 \approx 75\%$ of the ground state energy at $J = 0.45, h = 0.55$.
These results are comparable to those achieved recently by the Google team~\cite{mi2023stable}.

\section{Discussion}\label{sec:Discussion}


In this work, we introduced a dissipative variational protocol that emulates cooling a quantum system by coupling it to a bath. The protocol minimizes the steady-state energy achieved by repeatedly applying a cycle that consists of 
a block of unitary gates that act on 
the system and bath qubits
followed by a bath reset. Conceptually, it can be viewed as a variational counterpart of the simulated cooling approaches of Refs.~\onlinecite{Park2016, Kaplan2017, Mazzola2019, mathhies2022adibatic_demag, Piroli2024, Molpeceres2025, kishony2025gauged, kishony2025_chiral, mi2023stable, Lloyd2025, Ding2024, DingEndtoEnd, Zhan2025, Marti2025}, with the key advantage that it avoids 
the cost of simulating continuous time evolution
in the unitary part of the cycle. Instead, it employs an optimized shallow unitary circuit, making it particularly suitable for present-day NISQ devices.

A central ingredient is the optimization of the circuit parameters. In principle, this can be carried out using a hybrid classical-quantum approach, by iteratively measuring the steady-state energy on the quantum device and updating the parameters to reduce it. This approach has the benefit of directly targeting the system’s energy but suffers from the familiar drawback of requiring a large number of measurements to suppress shot noise in energy estimates. In this work, we chose instead to determine the parameters by classically simulating a small training system. Remarkably, for the transverse-field Ising model, parameters optimized on a system of $N=4$ spins performed well when applied to much larger systems, up to $N=28$. Whether such ``generalization'' from small to large systems holds for more challenging problems remains an open question.

More broadly, dissipative quantum algorithms are expected to have an advantage over purely unitary ground-state preparation methods (such as adiabatic state preparation or VQE) in cases where the target ground state cannot be easily connected to a simple product state. This situation arises when low-energy excitations are inherently non-local, as in topologically nontrivial states. The domain walls of the ferromagnetic Ising model provide the simplest example of such non-local excitations~\cite{mathhies2022adibatic_demag}. For such systems, the depth of unitary preparation circuits must grow with system size, since the energy gap above the ground state must close along the adiabatic path as the system grows. However, dissipative approaches also face difficulties when preparing systems with non-local excitations, because local couplings to the bath cannot efficiently remove single non-local excitations~\cite{mathhies2022adibatic_demag}. In some cases, non-local encodings of the system’s degrees of freedom can mitigate this limitation~\cite{kishony2025gauged,kishony2025_chiral}. Fermionic systems are a particularly compelling example, where coupling to a simulated fermionic bath can dramatically improve performance~\cite{kishony2025_chiral}. Extending the present variational cooling protocol to such fermionic problems is a natural and important next step.

Finally, an intriguing open problem is the preparation of thermal (Gibbs) states on NISQ devices. Since such states are inherently mixed, dissipative algorithms provide a natural framework, and several such protocols have been proposed~\cite{Rall2023, Chen2023, Chen2023EfficientGibbs, Ding2025, Guo2025, Hahn2025, Lloyd2025Thermal}. 
However, these typically require circuit depths far larger than those for ground-state preparation. Casting Gibbs-state preparation as a variational problem with a cost function that can be efficiently evaluated on a quantum device is nontrivial, and we leave its resolution to future work.





\section{Acknowledgments}
We thank N. H. Lindner for useful discussions, and A. Rosch for collaboration on closely related topics.
M.R. acknowledges the Brown Investigator Award, a program of the Brown Science Foundation, the University of Washington College of Arts and Sciences, and the Kenneth K. Young Memorial Professorship for support.
G.K. and E.B. were supported by CRC 183 of the Deutsche Forschungsgemeinschaft (subproject A01), a research grant from
the Estate of Gerald Alexander, ISF-MAFAT Quantum Science and Technology Grant no. 2478/24, and the CHE/PBC Doctoral Fellowships in Quantum Science and Technology 2024. 
We acknowledge the use of IBM Quantum Credits for this work. 
The views expressed are those of the authors, and do not reflect the official policy or position of IBM or the IBM Quantum team.
In this paper we used {\tt ibm\_kingston}, which is one of the IBM Heron R2 Quantum Processors.

\appendix
\section{Further notes on training}\label{app:training}

In the main text we discussed the primary considerations that go into training the Vari-Cool protocol for optimal performance.
As discussed there, it is important to train both for (i) a low expectation value of the steady state energy with respect to the target Hamiltonian, using a small number of layers $p$ per unitary block, and (ii) rapid cooling toward the steady state on a time scale of as few cycles as possible.
In this appendix we provide the final trained parameters used for the MPS simulations and experiments discussed in the main text, as well as a further discussion of training approaches. 

\subsection{Final trained parameters for simulations and experiments}
\begin{table}[t!]
\centering
\caption{Optimized parameters for $J=0.4$, $h=0.6$}
\label{tab:parameters_J0.4_h0.6}
\begin{tabular}{|c|c|c|c|c|}
\hline
Layer $\ell$ & $\alpha_\ell$ & $\beta_\ell$ & $\gamma_\ell$ & $\delta_\ell$ \\
\hline
1 & 0.010773 & -0.350834 &  3.141593 & 1.463996 \\
2 & 0.235103 & 1.264952 & 1.904127 & 1.062269 \\
3 & 0.566538 & 0.581135 & -0.765569 & 0.450404 \\
\hline
\end{tabular}
\end{table}
\begin{table}[t!]
\centering
\caption{Optimized parameters for $J=0.45$, $h=0.55$}
\label{tab:parameters_J0.45_h0.55}
\begin{tabular}{|c|c|c|c|c|}
\hline
Layer $\ell$ & $\alpha_\ell$ & $\beta_\ell$ & $\gamma_\ell$ & $\delta_\ell$ \\
\hline
1 & 0.116058 & -0.257237 & 3.141593 & 1.283935 \\
2 & 0.287760 & 1.238377 & 1.920788 & 0.940986 \\
3 & 0.720036 & 0.602624 & -0.830584 & 0.415904 \\
\hline
\end{tabular}
\end{table}
\begin{table}[t!]
\centering
\caption{Optimized parameters for $J=0.55$, $h=0.45$}
\label{tab:parameters_J0.55_h0.45}
\begin{tabular}{|c|c|c|c|c|}
\hline
Layer $\ell$ & $\alpha_\ell$ & $\beta_\ell$ & $\gamma_\ell$ & $\delta_\ell$ \\
\hline
1 & 0.073705 & -0.294230 & 3.141593 & 1.199571 \\
2 & 0.327663 & 1.102176 & 2.039226 & 0.991791 \\
3 & 0.865934 & 0.514179 & -0.687047 & 0.287476 \\
\hline
\end{tabular}
\end{table}


\begin{table}[t!]
\centering
\caption{Optimized parameters for $J=0.6$, $h=0.4$}
\label{tab:parameters_J0.6_h0.4}
\begin{tabular}{|c|c|c|c|c|}
\hline
Layer $\ell$ & $\alpha_\ell$ & $\beta_\ell$ & $\gamma_\ell$ & $\delta_\ell$ \\
\hline
1 & 0.165519 & -0.644517 & 3.141593 & 1.196658 \\
2 & 0.235756 & 1.256637 & 2.037732 & 0.724290 \\
3 & 0.935498 & 0.471761 & -0.876572 & 0.394391 \\
\hline
\end{tabular}
\end{table}
As discussed in the main text, we trained the protocol with $p = 3$ layers per unitary block using classical exact evolution with $N = 4$ system qubits and $n_{\rm bath} = 2$ bath qubits.
Using the training approaches described in the main text, we obtained the parameter sets in Tables \ref{tab:parameters_J0.4_h0.6} to \ref{tab:parameters_J0.6_h0.4} to use in the classical simulations (Sec.~\ref{sec:MPS}) and experimental demonstrations (Sec.~\ref{sec:QPU}) with $p = 3$ (see also Tables \ref{tab:parameters_J0.4_h0.6_p=4} to \ref{tab:parameters_J0.45_p=4} for parameters used in experimental tests with $p = 4$).
We performed the initial training for $J = 0.4, h = 0.6$ (Table \ref{tab:parameters_J0.4_h0.6}).
We used the optimized parameters at $J = 0.4, h = 0.6$ at the target depth $p = 3$ as the initial guess for the parameters at $J = 0.45, h = 0.55$, bootstrapping in this way up to $J = 0.6, h = 0.4$.
For the runs with $p = 4$ (see Sec.~\ref{subsec:ExtraData} and Fig.~\ref{fig:Results_6+3_p=4}), we used the pruning method to obtain optimized parameters at $J = 0.4, h = 0.6$.
We then used that parameter set as the initial guess at  $J = 0.45, h = 0.55$ with $p = 4$.
Note that all parameters were bounded to the range $[-\pi, \pi]$ during optimization.





\begin{figure*}[ht!]
    \centering
     \includegraphics[width=\textwidth]{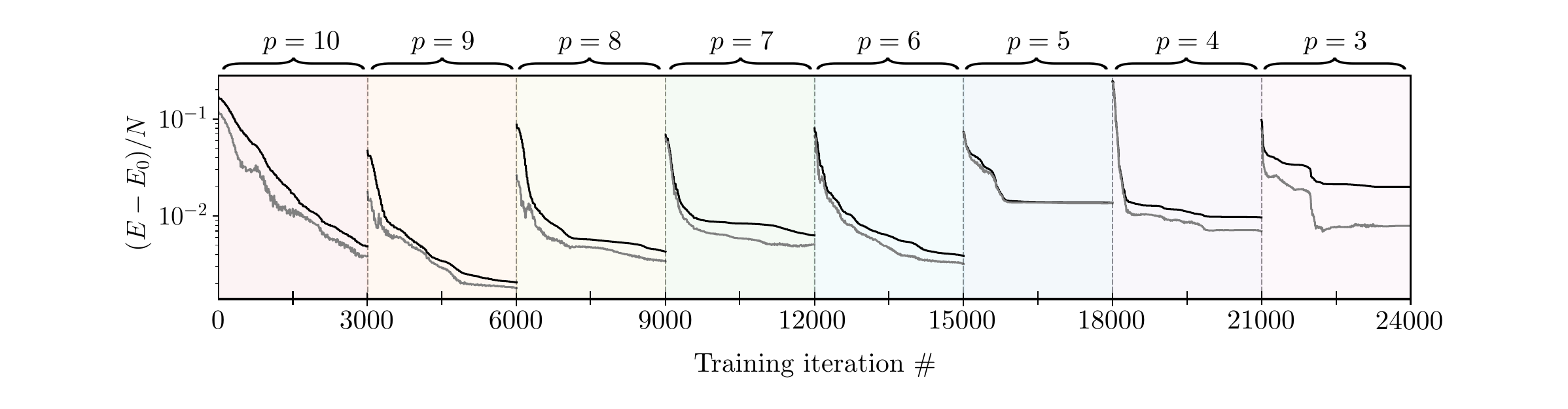}
    \caption{The layer-pruning method for training the Vari-Cool protocol.
    We begin with $p = 10$ layers per reset cycle.
    Within each training epoch with $p$ fixed we search for parameters to minimize the expectation value of the energy after 7 cycles using the Nelder-Mead optimizer.
    We then successively remove layers  
    and reoptimize over subsequent training epochs, down to a final circuit with $p = 3$ layers per reset cycle.
    The black solid line shows the expectation value of the energy, $E$, relative to the ground state energy, $E_0$, 
    after $T_{\rm train} = 7$ cycles; this value is used for the optimization in each training iteration.
    The gray line (below the black line) shows the steady state energy achieved by applying the circuit 40 times with the current values of the parameters at each training iteration.
    With $p = 3$ layers per cycle the steady state energy is close to the ground state energy.
    }
    \label{fig:pruning}
\end{figure*}
\subsection{Hyperparameters and training approaches}

The goal of the training process is to optimize the variational parameters in the protocol circuit to give low steady state energies with rapid cooling times, as described above.
In addition to the variational parameters $\boldsymbol{\alpha}, \boldsymbol{\beta}, \boldsymbol{\gamma}$, and $\boldsymbol{\delta}$, the protocol and the training process possess several {\it hyperparameters} that can be fixed or varied throughout the training:
\begin{itemize}
    \item The number of layers per unitary block, $p$ 
    \item The system size $N$ used during the training
    \item Hamiltonian parameters ($J$ and $h$ for the TFIM)
    \item The initial state $\Ket{\psi_0}$ supplied to the circuit
    \item The training time $T_\text{train}$ at which the energy is monitored for minimization
    \item The window size $\tau$ used for imposing the monotonicity constraint  
\end{itemize}

We have explored the roles of some of these hyperparameters in the training process.
As with many optimization problems, the initial guess for the variables being optimized over can also affect convergence and final results (especially in situations with many local minima).
We have also explored training using random initializations for the circuit parameters or parameter sets derived from Trotterization of continuous time cooling protocols.
In this appendix we share our observations on the roles of the hyperparameters and different initialization schemes to help guide future investigations and/or applications of the variational cooling protocol.

\subsubsection{Training with fixed $p$}
The simplest approach is to fix the number of layers per unitary block to the desired depth $p_*$, pick values for $T_{\rm train}$ and $\tau$, and set the system size and Hamiltonian parameters to those of the target system. 
We found that any particular guess for the initial values of the variational parameters will typically yield steady state (non-oscillating) behavior in the large cycle number limit.
For relatively large $p \gtrsim 7$, we found that obtaining initial guesses for the circuit parameters from Trotterized adiabatic demagnetization~\cite{mathhies2022adibatic_demag} provided a good starting point for optimization, with a low steady state energy.
However, for random parameter values or if Trotterization is used with a very small number of layers, $p$, the average energy in the steady state (before any optimization) will typically be high.

Due to the monotonically decreasing energy constraint, which we impose to avoid unsteady long-time behavior, it is important to choose an initial system state $\Ket{\psi_0}$ for the training with a {\it larger} energy expectation value than that of the steady state with the initially guessed parameters.
(Otherwise, the energy will {\it increase} after the cycle is applied, and optimization will not be able to proceed.)
By choosing an initial system state $\Ket{\psi_0}$ with energy larger than that of the initial (un-optimized) steady state, the optimizer will be able to begin its work of driving the steady state energy down.
Note that this situation can also be avoided by selectively applying the the monotonicity constraint only if non-steady long time behavior is being generated.

For the training displayed in Fig.~\ref{fig:Training} of the main text, we fixed an initial state $\Ket{\psi_0} = \Ket{0000}$, and tested 15,000 sets of initial guesses~\footnote{We note that initial guesses populated with many $0$ values do not give much cooling, and thus typically do not provide a good starting point for optimization (cf.~Ref.~\onlinecite{Cerezo2021}).} for the circuit parameters (there we used $p = 3$, for a total of 12 variational parameters in the circuit). 
We screened the parameter sets by checking for monotonically decreasing energy starting from the state $\Ket{\psi_0}$, and selected the 10 best sets with the lowest steady state energies for further optimization.
We then ran the training with $T_{\rm train} = 7$, $\tau = 2$ for each of those initial guesses, and picked the final parameter set that gave the lowest steady state energy.
This approach is straightforward, easy to implement, and robustly gave steady states with low energies relative to the ground state.

In some cases we observed that the steady state energy could be driven even lower by varying hyperparameters such as $T_{\rm train}$ or the initial state $\Ket{\psi_0}$ and then re-training.
We found that multiple rounds of hyperparameter variation and re-optimization could also be helpful in some cases.
For a well chosen initial state $\Ket{\psi_0}$ and good coverage over the space of possible parameter values, one round of training without any further tuning of hyperparameters was often enough to obtain acceptable results.

\subsubsection{The layer-pruning training method}
The ``pruning'' method offers a straightforward and consistent way to achieve good performance for small numbers of layers $p$ per unitary block by systematically reducing the value of $p$ between training epochs.
The idea is to start with a moderate value of $p$, where naive Trotterization of a simulated cooling protocol gives a good starting point for optimization.
Then, after optimizing the circuit parameters at fixed $p$, we successively reduce the number of layers and re-optimize, reducing the value of $p$ step-by-step down to the desired value, $p_*$.

An example training run using the layer-pruning method is shown in Fig.~\ref{fig:pruning}.
Here we start with $p = 10$ layers per unitary block.
For the initial circuit parameters, we take a coarse 10-step (first-order) Trotterization of the simulated adiabatic demagnetization protocol in Ref.~\onlinecite{mathhies2022adibatic_demag}.
Define $t_\ell = \ell \Delta t$, where $\ell$ is the layer index and $\Delta t = T/p$ is the Trotter time step, with $T$ denoting the sweep time of the simulated cooling cycle.
For the example in Fig.~\ref{fig:pruning} we used $T = 5$. 
We then set
$\alpha_\ell = 2J\Delta t$,  $\beta_\ell = 2h\Delta t$, $\gamma_\ell = 2 g(t_\ell) \Delta t$, $\delta_{\ell}=2 B(t_\ell)\Delta t$,
where $g(t)$ and $B(t)$,  describe the ramps of the system-bath coupling and bath Zeeman field, respectively.
We use the smooth functional forms for $g(t)$ and $B(t)$ as given in the supplementary material of Ref.~\onlinecite{kishony2025gauged}.
We furthermore take the initial state of the system qubits to be $\Ket{0000}$.

As shown in Fig.~\ref{fig:pruning}, the energy at $T_{\rm train}$ (black lines) 
decreases with successive training iterations within each training epoch of fixed $p$. 
The corresponding steady state energy (gray lines, below the black lines) also tends to decrease, though not monotonically throughout each epoch.
Each time one layer of parameters is removed, the energy jumps up discontinuously.
For $p > 7$, we pruned the circuit by removing one whole layer; we checked the value of the steady state energy obtained after removing each of the $p$ possible choices for which layer to prune, and chose the one that left the lowest steady state energy (smallest jump) after removal.
For $p \le 7$, we pruned the circuit by checking all possible combinations of removing one of the $\alpha_\ell$'s, one of the $\beta_\ell$'s, one of the $\gamma_\ell$'s, and one of the $\delta_\ell$'s, and choosing the smallest jump of the steady state energy after removal.
Further optimization within each epoch of fixed $p$ drove the steady state energy back down.
As the figure shows, the steady state energy at $p=3$ is only moderately higher than its value for much a larger depth circuit with $p = 10$.
Further optimization is possible by retraining with different initial states, and/or by varying the value of $T_{\rm train}$.\\


\subsubsection{The role of $T_{\rm train}$ and the monotonicity constraint}
As mentioned in the main text, to achieve rapid cooling it is important to train the protocol using the energy expectation value after a limited number of cycles, $T_{\rm train}$.
In this section we discuss important further aspects of training at small values of $T_{\rm train}$.

For concreteness, consider the strategy of minimizing the energy after one cycle, either for a fixed or a random initial state.
Notice that the input to the second cycle is the output from the first; unless the initial state happens to be the steady state $\hat{\rho}_{\rm steady}$, the second cycle acts on a {\it different} state $\hat{\rho}' = \mathcal{E}[\hat{\rho}]$ than the one one which the circuit was trained.
(Here, as in the main text, $\mathcal{E}$ is the quantum channel that describes one cycle of the protocol.)
The energy of the system could thus increase or decrease after the second cycle, depending on specifics of the circuit.

\begin{figure}[t]
    \centering
    \includegraphics[width=\linewidth]{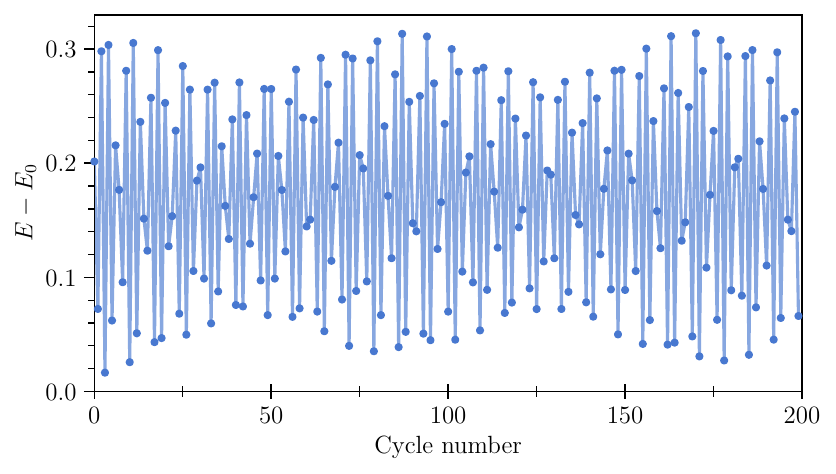}
    \caption{Example of unsteady, limit cycle behavior.
    If the Vari-Cool circuit is trained without imposing a constraint that the energy should decrease under repeated cycles, optimized parameters may yield low energies precisely at the training time $T_{\rm train}$ but unsteady behavior under further repeated applications of the cycle.
    In this situation, the channel does not have a well-defined steady state, and its behavior is sensitive to the initial state of the system qubits.
    For the example shown, we used $T_{\rm train} = 1$ with initial state $\Ket{++++}$, and $J = 0.4, h = 0.6$.
    Notice that the energy decreases over the first cycle, but then increases on the next, continuing to oscillate thereafter.
    }
    \label{fig:LimitCycle}
\end{figure}
Indeed, as illustrated in Fig.~\ref{fig:LimitCycle}, naive training based on the energy after a small number of cycles, $T_{\rm train}$, can lead to unsteady or limit cycle type behavior of the channel.
Specifically, for the initial state on which it was trained, the channel achieves a low energy following the application of precisely $T_{\rm train}$ cycles.
However, under repeated applications, the energy continues to fluctuate (periodically or quasiperiodically) and does not reach a steady state on the time scale examined here.

As discussed in the main text, such unsteady behavior is easy to avoid by imposing a {\it monotonicity} constraint on the optimization.
During training we only accept parameter sets for which the energy decreases monotonically in a window of cycles centered around $T_{\rm train}$, from $T_{\rm train} - \tau$ to $T_{\rm train} + \tau$. 
The search space is thus limited to circuits that give rise to stable steady states at long times.

Training for a low energy at $T_{\rm train}$ cycles, with the monotonicity constraint included, gives low steady state energies and short cooling times.
As is evident in Fig.~\ref{fig:Training}b in the main text, under repeated cycles the energy continues to drop past its value at $T_{\rm train}$.
We have not systematically explored the dependence of the cooling time and steady state energy on $T_{\rm train}$ and $\tau$, but for specific applications the protocol could in principle be further optimized with respect to these hyperparameters.

\begin{figure}[t]
    \centering
   \includegraphics[width=\linewidth]{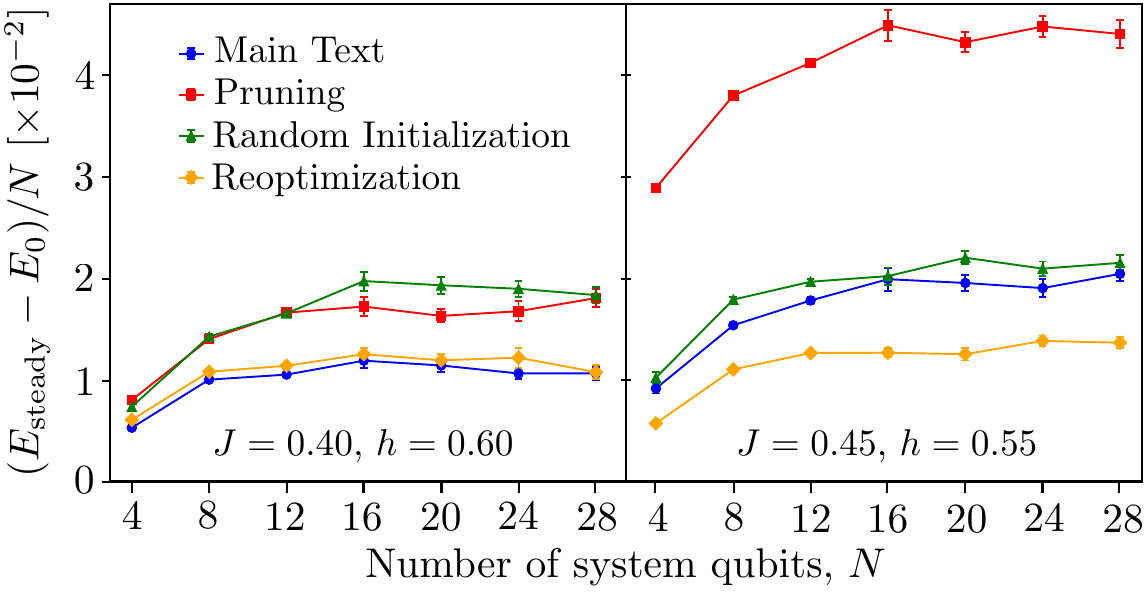}
    \caption{Transferability of protocol training from small to large system sizes.
    We trained the protocol on classical hardware via exact simulation with $N = 4$ system sites coupled to $n_{\rm bath} = 2$ bath sites for (left panel) $J = 0.4, h = 0.6$ and (right panel) $J = 0.45, h = 0.55$.
    The different colors/symbols show the residual steady state energy density vs.~system size from MPS-based simulations (without added noise) using four independently trained sets of parameters.
    The training methods used are indicated on the legend.
    The parameters for the blue circles are the same as those used in the main text.
    The parameters for the red squares were obtained via the ``layer pruning'' method.
    The parameters for the green triangles were obtained by optimization at fixed $p = 3$ starting from many random initial guesses and down-selecting.
    The parameters for the orange diamonds were obtained using a ``reoptimization'' strategy in which we varied the initial system state supplied to the circuit between training epochs at fixed $p = 3$ to help drive the steady state energy lower.
    For the reoptimization we removed the monotonicity constraint; this helped to further decrease the steady state energies. 
    The observed behavior is similar for all four parameter sets, showing a modest increase of the energy density with system size and rapid saturation by approximately $N = 12$.
    }
    \label{fig:Transferability}
\end{figure}
\subsection{Transferability to larger system sizes}
In this appendix we demonstrate the transferability of our translation-invariant ansatz (see Fig.~\ref{fig:Transferability}).
As noted in the main text, we trained the circuit parameters using simulated exact evolution of a small system ($N = 4$ system sites coupled to $n_{\rm bath} = 2$ bath sites) on a classical computer.
We then used the optimized parameters obtained from those simulations in the MPS-based simulations and experimental runs at sizes up to $N = 28$ system qubits coupled to $n_{\rm bath} = 14$ bath sites.

As discussed above, we found that a variety of methods can be used to obtain good parameters for the protocol.
To examine the robustness of the transferability to larger system sizes, we trained the model four separate times at two different points in the phase diagram ($J = 0.4, h = 0.6$ and $J = 0.45, h = 0.55$) to obtain distinct parameter sets that give comparable steady state energies at system size $N = 4$.
We then tracked the steady state energy density (without any added noise/errors) as a function of system size, up to $N = 28$ system qubits coupled to $n_{\rm bath} = 14$ bath sites, for each of these parameter sets, using MPS-based simulations.
As in the main text, we use the average value of the system energy after 40 cycles as a proxy for the steady state energy.

The data in Fig.~\ref{fig:Transferability} show that the steady state energy density increases slightly with system size, but in most cases saturates at a moderate level by about $N = 12$.
As expected, there are slight variations between the results obtained for the different parameter sets, but overall the behavior is consistent and robust to the training details.
Given the modest increases of the energy density with system size, we expect that these parameters obtained by rapid training at small system size would provide a good starting point for additional (much more costly) optimization at larger system size, if desired.
It is an interesting question for future work to what extent this behavior persists in more strongly correlated models.

\section{Further experimental data}\label{app:QPU}
\subsection{Calibration data and device layouts}
In this section we provide key calibration data for the device used in our experimental tests, and specify the particular qubits used for the runs with $N = 6$ system qubits.
The layout for the runs with $N = 28$ system qubits is shown in Fig.~\ref{fig:QPU} of the main text.

\begin{figure}[t]
    \centering
   \includegraphics[width=\linewidth]{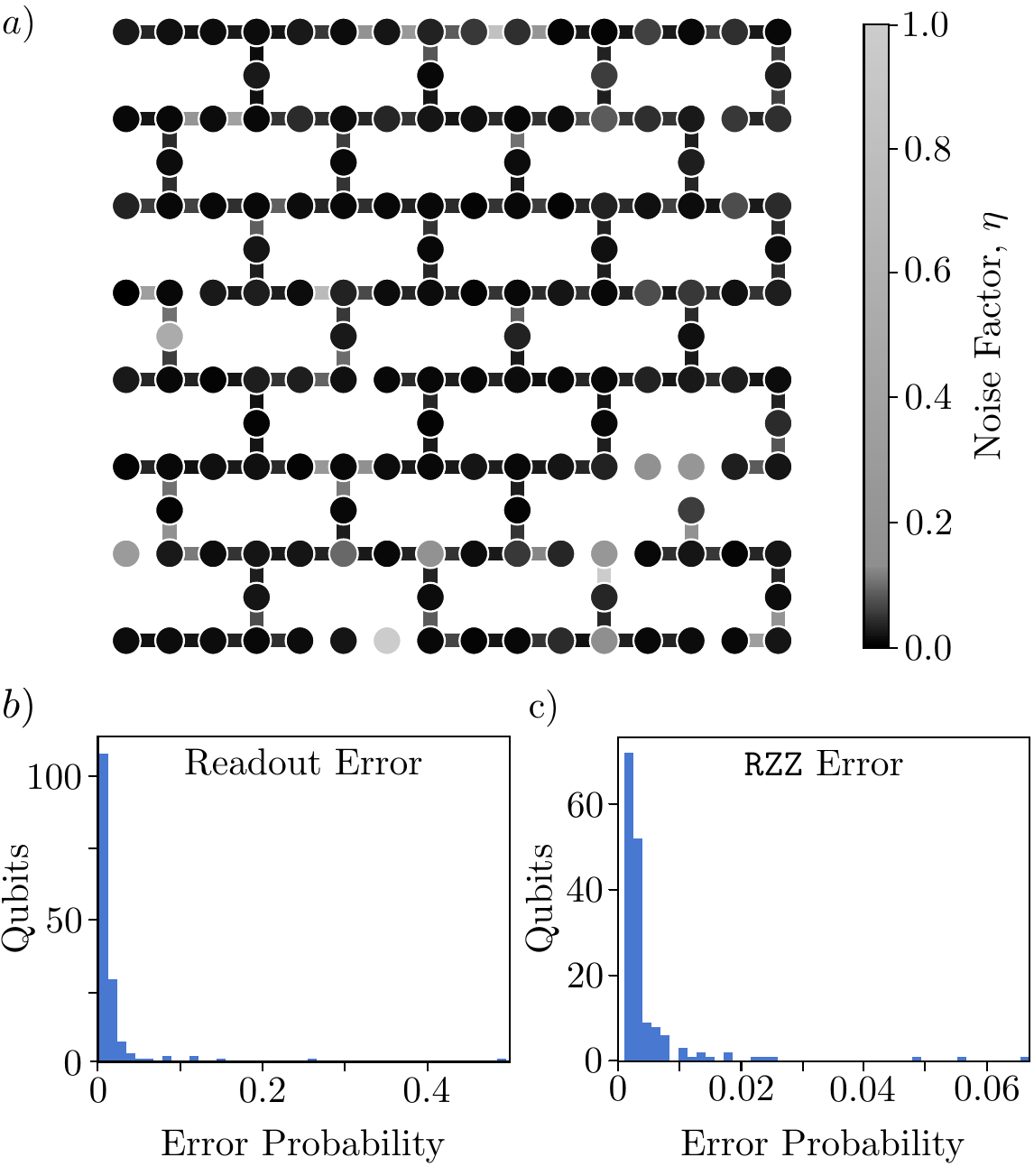}

    \caption{Qubit performance data. 
    a) Diagram of the {\tt ibm\_kingston} quantum processor showing calibration data for single qubit readout error and two qubit ({\tt RZZ}) gate error probabilities at the time our experiments were performed.
    The color scale ranges from $\eta = 0$ to $\eta = 1$, representing a multiplicative factor for the error probabilities: $p_{\rm readout} = \eta\, p^{\rm (max)}_{\rm readout}$ and $p_{\tt RZZ} = \eta\, p^{\rm (max)}_{\tt RZZ}$, with $p^{\rm (max)}_{\rm readout} = 0.497$ 
    and $p^{\rm (max)}_{\tt RZZ} = 0.0670$. 
    b) Histogram of qubit readout error probabilities. 
    c) Histogram of two-qubit {\tt RZZ} gate error probabilities.
    The large peaks in both histograms, as well as the fact that most nodes and links in panel (a) show up as black, indicate that a large fraction of the qubits possess similar, good performance. 
    We tailor the qubit layouts for our experiments to avoid the outlier qubits and bonds with large readout errors or {\tt RZZ} gate errors (see Figs.~\ref{fig:QPU} and \ref{fig:DeviceLayouts}). 
    }
    \label{fig:Calibration}
\end{figure}
All of our experimental data were collected on the {\tt ibm\_kingston} quantum processor.
We chose qubit layouts 
based on reported calibration data at the time of the runs.
Specifically, we aimed to use qubits with low single and two qubit gate errors, and high readout fidelities.
In Fig.~\ref{fig:Calibration} we display the qubit readout assignment error probability and the {\tt RZZ} gate error probability using grayscale intensity on the sites and bonds, respectively.
As written on IBM's website, ``Readout error is measured by preparing the qubit in the 0 (1) state and measuring the probability of an output in the 1 (0) state. The reported value is the average of these two errors. The median is taken over all qubits,'' while the {\tt RZZ} gate error represents the ``error in the {\tt RZZ} gate averaged over the {\tt RZZ} angles using a variant of randomized benchmarking for arbitrary unitaries''~\cite{IBM_Quantum_ErrorDefs}.

\begin{figure}[t]
    \centering
    \includegraphics[width=\linewidth]{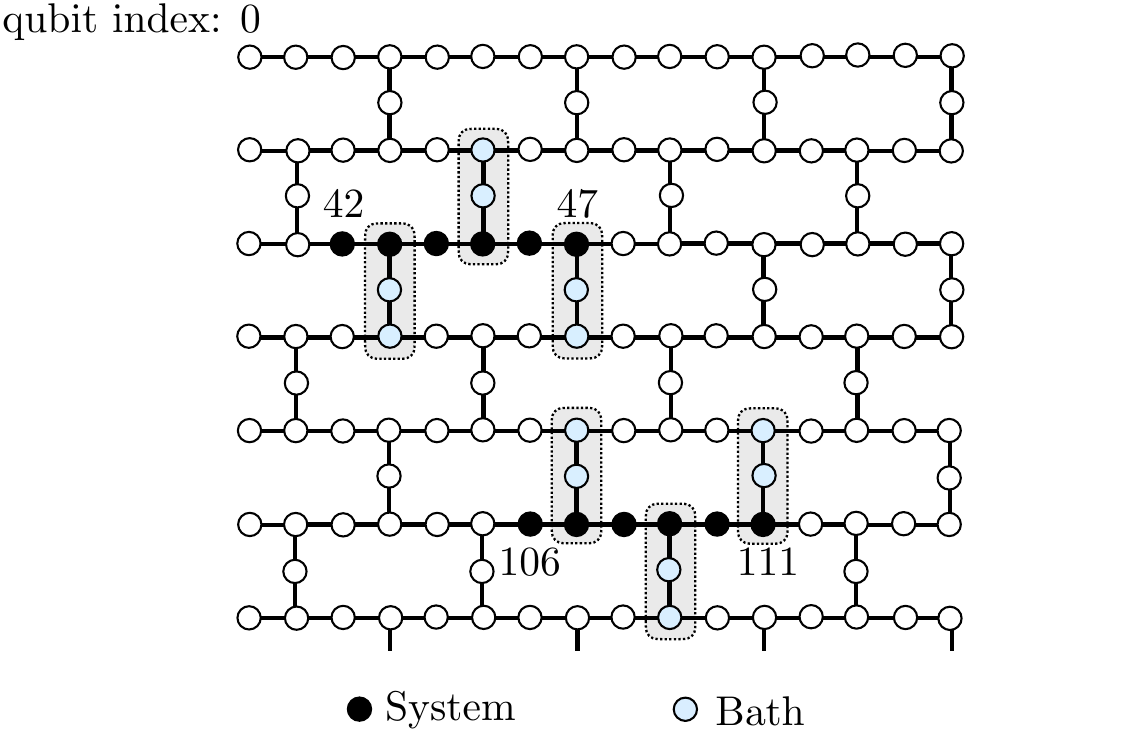}
    \caption{For system size $N = 6$ with $n_{\rm bath} = 3$ bath sites, we demonstrated the Vari-Cool protocol experimentally on the two disjoint sets of qubits shown above.
    The results shown in the main text were obtained from the upper set of qubits (with the system chain running from qubit 42 to qubit 47).
    Results for the lower chain (running from qubit 106 to qubit 111) are shown in Fig.~\ref{fig:Results_6+3_second_geom}.}
    \label{fig:DeviceLayouts}
\end{figure}

For system size $N = 6$ with $n_{\rm bath} = 3$ bath sites, we ran the protocol in parallel on two disjoint sets of qubits on the {\tt ibm\_kingston} quantum processor.
The specific qubit layouts that we used are shown in Fig.~\ref{fig:DeviceLayouts}.
See Ref.~\onlinecite{IBM_Quantum} for further details of the {\tt ibm\_kingston} processor.

\subsection{Additional data for $N = 6$, $n_{\rm bath}$ = 3}
\label{subsec:ExtraData}
\begin{figure}[t]
    \centering
   \includegraphics[width=\linewidth]{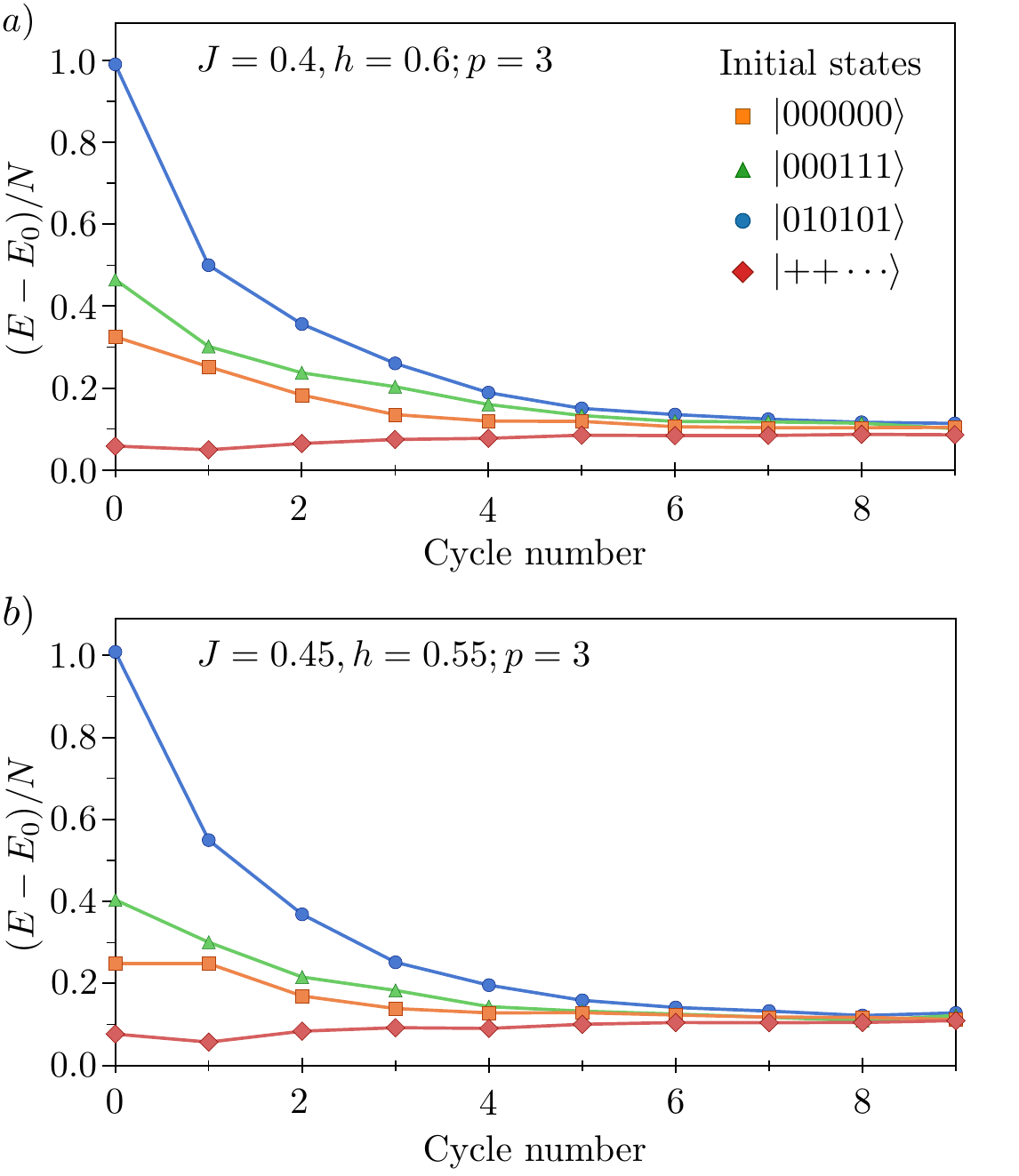}
    \caption{Experimental demonstration of the cooling algorithm on $N= 6$ system sites, obtained in parallel (on a disjoint set of qubits on the same chip) with the data in Fig.~\ref{fig:Results_6+3}. The data shown here were obtained using the system qubits 106 to 111 in Fig. \ref{fig:DeviceLayouts}. 
    Data were obtained using 8192 shots for each set of mutually compatible observables.
    a) Results for $J=0.4, h = 0.6$, with ground state energy $E_0 = -3.9390$. 
    The average residual energy above the ground state is approximately
    $0.101 \pm 0.010$ per system site.
    b) Results for $J = 0.45, h = 0.55$, with ground state energy $E_0 = -3.7720$. 
     The average residual energy above the ground state is approximately $0.118 \pm 0.008$ per system site.
    }
\label{fig:Results_6+3_second_geom}
\end{figure}

Here we present additional results for system size $N = 6$, with $n_{\rm bath} = 3$ bath sites to complement the experimental data shown in Fig.~\ref{fig:Results_6+3} of the main text.
Figure \ref{fig:Results_6+3_second_geom} shows results on the lower set of qubits shown on the device layout in Fig.~\ref{fig:DeviceLayouts}, obtained with the same parameters and in parallel with the data for Fig.~\ref{fig:Results_6+3}.
\begin{figure}[t]
    \centering
   \includegraphics[width=\linewidth]{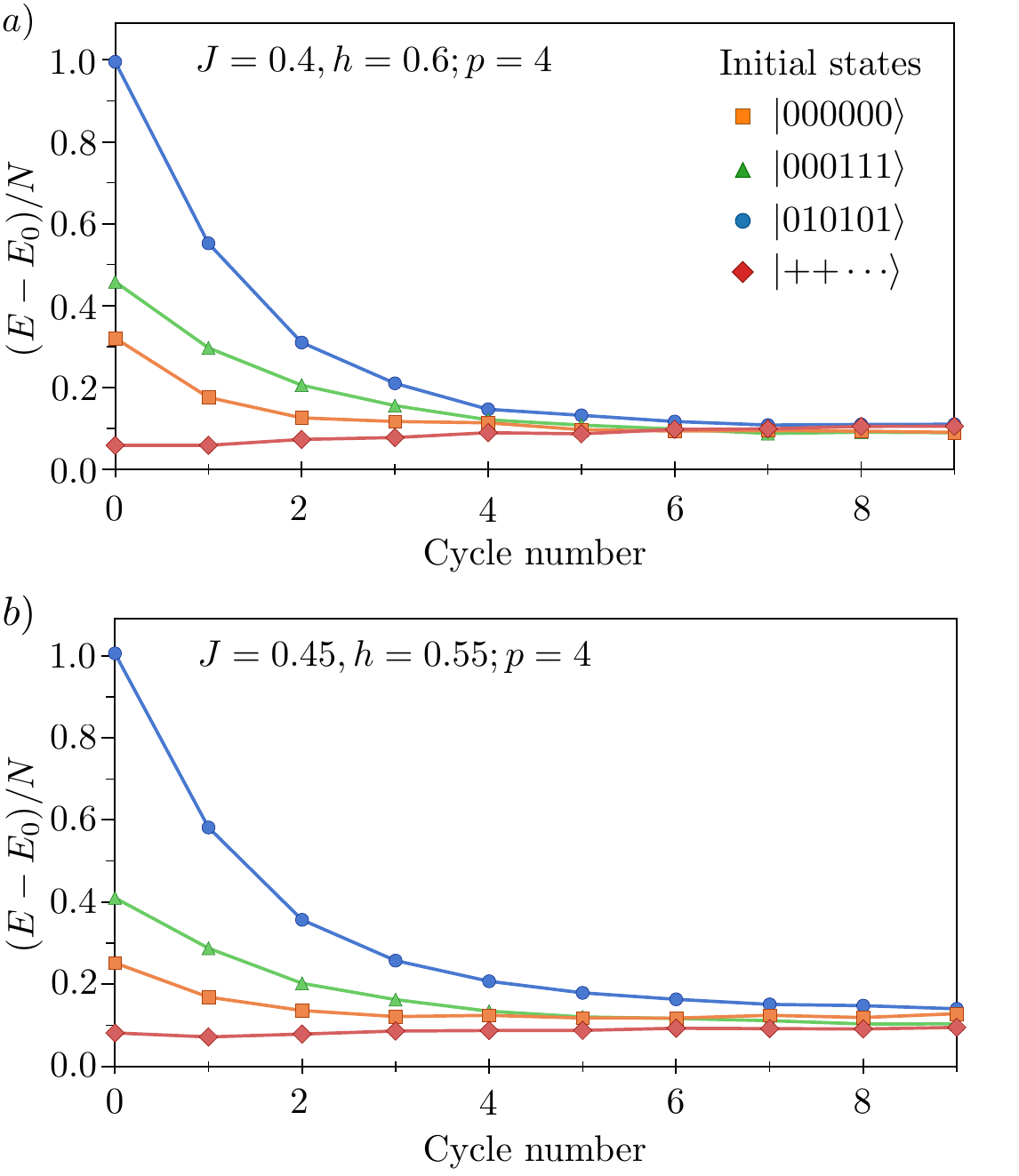}
    \caption{Experimental demonstration of the cooling algorithm using $p = 4$ layers per unitary block, on $N = 6$ system sites with $n_{\rm bath} = 3$ bath sites.
    Here we used the upper chain of qubits shown in Fig.~\ref{fig:DeviceLayouts}.
    Data were obtained using 8192 shots for each measurement basis ($Z$ or $X$).
    a) Results for $J=0.4, h = 0.6$, with ground state energy $E_0 = -3.9390$. 
    The average residual energy above the ground state in the steady state is 
    $0.0988 \pm 0.009$ per system site.
    b) Results for $J = 0.45, h = 0.55$, with ground state energy $E_0 = -3.7720$. 
     The average residual energy above the ground state in the steady state is
     $0.117 \pm 0.017$ per system site.
    }
\label{fig:Results_6+3_p=4}
\end{figure}

Comparing Fig.~\ref{fig:Results_6+3_second_geom} with Fig.~\ref{fig:Results_6+3} in the main text, we see that the protocol performs similarly on the two sets of qubits.
We observe slight differences in the curves and steady state values, as expected due to variations in qubit parameters and environmental noise across the device. 

In Fig.~\ref{fig:Results_6+3_p=4} we show results using $p = 4$ layers per unitary block.
For these runs we used the trained parameters shown in Tables \ref{tab:parameters_J0.4_h0.6_p=4} and \ref{tab:parameters_J0.45_p=4}.
Although in the noiseless simulations used for training the circuit with $p = 4$ layers per unitary block is able to reach lower steady state residual energies than that with $p = 3$, the experimental results show no significant differences between $p = 4$ and $p = 3$ (compare Fig.~\ref{fig:Results_6+3_p=4} with Figs.~\ref{fig:Results_6+3} and \ref{fig:Results_6+3_second_geom}).
In particular, we see a somewhat larger spread of final energies (after 9 cycles) at $J = 0.45, h = 0.55$ with $p = 4$, which may reflect a greater influence of device noise in this case.
However, the average final energy at 9 cycles is only slightly raised compared to that in the $p = 3$ case.
\newpage


\begin{table}[b!]
\centering
\caption{Optimized parameters for $J=0.4$, $h=0.6$ ($p=4$)}
\label{tab:parameters_J0.4_h0.6_p=4}
\begin{tabular}{|c|c|c|c|c|}
\hline
Layer $\ell$ & $\alpha_\ell$ & $\beta_\ell$ & $\gamma_\ell$ & $\delta_\ell$ \\
\hline
1 & 0.048970 & -0.293862 & -3.141593 & 1.971756 \\
2  & 0.037290 & 1.096104  & 3.141593  & 1.176287 \\
3 & 0.460928 & 0.708252  & -2.229250 & 1.399117 \\
4 & 0.523253 & 0.457410  & -1.986812 & -0.487790 \\
\hline
\end{tabular}
\end{table}
\begin{table}[b!]
\centering
\caption{Optimized parameters for $J=0.45$, ($p=4$)}
\label{tab:parameters_J0.45_p=4}
\begin{tabular}{|c|c|c|c|c|}
\hline
Layer $\ell$ & $\alpha_\ell$ & $\beta_\ell$ & $\gamma_\ell$ & $\delta_\ell$ \\
\hline
1 & 0.055091 & -0.269373 & -3.141593 & 1.971756 \\
2  & 0.041951 & 1.004762  & 3.141593  & 1.176287 \\
3 & 0.518544 & 0.649231  & -2.229250 & 1.399117 \\
4 & 0.588660 & 0.419293  & -1.986812 & -0.487790 \\
\hline
\end{tabular}
\end{table}

\clearpage

\bibliography{References}

\begin{thebibliography}{51}%
\makeatletter
\providecommand \@ifxundefined [1]{%
 \@ifx{#1\undefined}
}%
\providecommand \@ifnum [1]{%
 \ifnum #1\expandafter \@firstoftwo
 \else \expandafter \@secondoftwo
 \fi
}%
\providecommand \@ifx [1]{%
 \ifx #1\expandafter \@firstoftwo
 \else \expandafter \@secondoftwo
 \fi
}%
\providecommand \natexlab [1]{#1}%
\providecommand \enquote  [1]{``#1''}%
\providecommand \bibnamefont  [1]{#1}%
\providecommand \bibfnamefont [1]{#1}%
\providecommand \citenamefont [1]{#1}%
\providecommand \href@noop [0]{\@secondoftwo}%
\providecommand \href [0]{\begingroup \@sanitize@url \@href}%
\providecommand \@href[1]{\@@startlink{#1}\@@href}%
\providecommand \@@href[1]{\endgroup#1\@@endlink}%
\providecommand \@sanitize@url [0]{\catcode `\\12\catcode `\$12\catcode `\&12\catcode `\#12\catcode `\^12\catcode `\_12\catcode `\%12\relax}%
\providecommand \@@startlink[1]{}%
\providecommand \@@endlink[0]{}%
\providecommand \url  [0]{\begingroup\@sanitize@url \@url }%
\providecommand \@url [1]{\endgroup\@href {#1}{\urlprefix }}%
\providecommand \urlprefix  [0]{URL }%
\providecommand \Eprint [0]{\href }%
\providecommand \doibase [0]{https://doi.org/}%
\providecommand \selectlanguage [0]{\@gobble}%
\providecommand \bibinfo  [0]{\@secondoftwo}%
\providecommand \bibfield  [0]{\@secondoftwo}%
\providecommand \translation [1]{[#1]}%
\providecommand \BibitemOpen [0]{}%
\providecommand \bibitemStop [0]{}%
\providecommand \bibitemNoStop [0]{.\EOS\space}%
\providecommand \EOS [0]{\spacefactor3000\relax}%
\providecommand \BibitemShut  [1]{\csname bibitem#1\endcsname}%
\let\auto@bib@innerbib\@empty
\bibitem [{\citenamefont {{Cirac}}\ and\ \citenamefont {{Zoller}}(2012)}]{2012CiracZoller}%
  \BibitemOpen
  \bibfield  {author} {\bibinfo {author} {\bibfnamefont {J.~I.}\ \bibnamefont {{Cirac}}}\ and\ \bibinfo {author} {\bibfnamefont {P.}~\bibnamefont {{Zoller}}},\ }\bibfield  {title} {\bibinfo {title} {{Goals and opportunities in quantum simulation}},\ }\href {https://doi.org/10.1038/nphys2275} {\bibfield  {journal} {\bibinfo  {journal} {Nature Physics}\ }\textbf {\bibinfo {volume} {8}},\ \bibinfo {pages} {264} (\bibinfo {year} {2012})}\BibitemShut {NoStop}%
\bibitem [{\citenamefont {Georgescu}\ \emph {et~al.}(2014)\citenamefont {Georgescu}, \citenamefont {Ashhab},\ and\ \citenamefont {Nori}}]{Georgescu_2014}%
  \BibitemOpen
  \bibfield  {author} {\bibinfo {author} {\bibfnamefont {I.}~\bibnamefont {Georgescu}}, \bibinfo {author} {\bibfnamefont {S.}~\bibnamefont {Ashhab}},\ and\ \bibinfo {author} {\bibfnamefont {F.}~\bibnamefont {Nori}},\ }\bibfield  {title} {\bibinfo {title} {Quantum simulation},\ }\href {https://doi.org/10.1103/revmodphys.86.153} {\bibfield  {journal} {\bibinfo  {journal} {Reviews of Modern Physics}\ }\textbf {\bibinfo {volume} {86}},\ \bibinfo {pages} {153} (\bibinfo {year} {2014})}\BibitemShut {NoStop}%
\bibitem [{\citenamefont {Zhou}\ \emph {et~al.}(2020)\citenamefont {Zhou}, \citenamefont {Wang}, \citenamefont {Choi}, \citenamefont {Pichler},\ and\ \citenamefont {Lukin}}]{ZhouQAOAReview}%
  \BibitemOpen
  \bibfield  {author} {\bibinfo {author} {\bibfnamefont {L.}~\bibnamefont {Zhou}}, \bibinfo {author} {\bibfnamefont {S.-T.}\ \bibnamefont {Wang}}, \bibinfo {author} {\bibfnamefont {S.}~\bibnamefont {Choi}}, \bibinfo {author} {\bibfnamefont {H.}~\bibnamefont {Pichler}},\ and\ \bibinfo {author} {\bibfnamefont {M.~D.}\ \bibnamefont {Lukin}},\ }\bibfield  {title} {\bibinfo {title} {{Quantum Approximate Optimization Algorithm: Performance, Mechanism, and Implementation on Near-Term Devices}},\ }\href {https://doi.org/10.1103/PhysRevX.10.021067} {\bibfield  {journal} {\bibinfo  {journal} {Phys. Rev. X}\ }\textbf {\bibinfo {volume} {10}},\ \bibinfo {pages} {021067} (\bibinfo {year} {2020})}\BibitemShut {NoStop}%
\bibitem [{\citenamefont {Moll}\ \emph {et~al.}(2018)\citenamefont {Moll}, \citenamefont {Barkoutsos}, \citenamefont {Bishop}, \citenamefont {Chow}, \citenamefont {Cross}, \citenamefont {Egger}, \citenamefont {Filipp}, \citenamefont {Fuhrer}, \citenamefont {Gambetta}, \citenamefont {Ganzhorn}, \citenamefont {Kandala}, \citenamefont {Mezzacapo}, \citenamefont {Müller}, \citenamefont {Riess}, \citenamefont {Salis}, \citenamefont {Smolin}, \citenamefont {Tavernelli},\ and\ \citenamefont {Temme}}]{Moll_2018}%
  \BibitemOpen
  \bibfield  {author} {\bibinfo {author} {\bibfnamefont {N.}~\bibnamefont {Moll}}, \bibinfo {author} {\bibfnamefont {P.}~\bibnamefont {Barkoutsos}}, \bibinfo {author} {\bibfnamefont {L.~S.}\ \bibnamefont {Bishop}}, \bibinfo {author} {\bibfnamefont {J.~M.}\ \bibnamefont {Chow}}, \bibinfo {author} {\bibfnamefont {A.}~\bibnamefont {Cross}}, \bibinfo {author} {\bibfnamefont {D.~J.}\ \bibnamefont {Egger}}, \bibinfo {author} {\bibfnamefont {S.}~\bibnamefont {Filipp}}, \bibinfo {author} {\bibfnamefont {A.}~\bibnamefont {Fuhrer}}, \bibinfo {author} {\bibfnamefont {J.~M.}\ \bibnamefont {Gambetta}}, \bibinfo {author} {\bibfnamefont {M.}~\bibnamefont {Ganzhorn}}, \bibinfo {author} {\bibfnamefont {A.}~\bibnamefont {Kandala}}, \bibinfo {author} {\bibfnamefont {A.}~\bibnamefont {Mezzacapo}}, \bibinfo {author} {\bibfnamefont {P.}~\bibnamefont {Müller}}, \bibinfo {author} {\bibfnamefont {W.}~\bibnamefont {Riess}}, \bibinfo {author} {\bibfnamefont {G.}~\bibnamefont {Salis}}, \bibinfo {author} {\bibfnamefont {J.}~\bibnamefont
  {Smolin}}, \bibinfo {author} {\bibfnamefont {I.}~\bibnamefont {Tavernelli}},\ and\ \bibinfo {author} {\bibfnamefont {K.}~\bibnamefont {Temme}},\ }\bibfield  {title} {\bibinfo {title} {Quantum optimization using variational algorithms on near-term quantum devices},\ }\href {https://doi.org/10.1088/2058-9565/aab822} {\bibfield  {journal} {\bibinfo  {journal} {Quantum Science and Technology}\ }\textbf {\bibinfo {volume} {3}},\ \bibinfo {pages} {030503} (\bibinfo {year} {2018})}\BibitemShut {NoStop}%
\bibitem [{\citenamefont {Lanyon}\ \emph {et~al.}(2010)\citenamefont {Lanyon}, \citenamefont {Whitfield}, \citenamefont {Gillett}, \citenamefont {Goggin}, \citenamefont {Almeida}, \citenamefont {Kassal}, \citenamefont {Biamonte}, \citenamefont {Mohseni}, \citenamefont {Powell}, \citenamefont {Barbieri}, \citenamefont {Aspuru-Guzik},\ and\ \citenamefont {White}}]{LanyonQCReview}%
  \BibitemOpen
  \bibfield  {author} {\bibinfo {author} {\bibfnamefont {B.~P.}\ \bibnamefont {Lanyon}}, \bibinfo {author} {\bibfnamefont {J.~D.}\ \bibnamefont {Whitfield}}, \bibinfo {author} {\bibfnamefont {G.~G.}\ \bibnamefont {Gillett}}, \bibinfo {author} {\bibfnamefont {M.~E.}\ \bibnamefont {Goggin}}, \bibinfo {author} {\bibfnamefont {M.~P.}\ \bibnamefont {Almeida}}, \bibinfo {author} {\bibfnamefont {I.}~\bibnamefont {Kassal}}, \bibinfo {author} {\bibfnamefont {J.~D.}\ \bibnamefont {Biamonte}}, \bibinfo {author} {\bibfnamefont {M.}~\bibnamefont {Mohseni}}, \bibinfo {author} {\bibfnamefont {B.~J.}\ \bibnamefont {Powell}}, \bibinfo {author} {\bibfnamefont {M.}~\bibnamefont {Barbieri}}, \bibinfo {author} {\bibfnamefont {A.}~\bibnamefont {Aspuru-Guzik}},\ and\ \bibinfo {author} {\bibfnamefont {A.~G.}\ \bibnamefont {White}},\ }\bibfield  {title} {\bibinfo {title} {{Towards quantum chemistry on a quantum computer}},\ }\href@noop {} {\bibfield  {journal} {\bibinfo  {journal} {Nature Chemistry}\ }\textbf {\bibinfo {volume} {2}}
  (\bibinfo {year} {2010})}\BibitemShut {NoStop}%
\bibitem [{\citenamefont {Bauer}\ \emph {et~al.}(2020)\citenamefont {Bauer}, \citenamefont {Bravyi}, \citenamefont {Motta},\ and\ \citenamefont {Chan}}]{BauerQC}%
  \BibitemOpen
  \bibfield  {author} {\bibinfo {author} {\bibfnamefont {B.}~\bibnamefont {Bauer}}, \bibinfo {author} {\bibfnamefont {S.}~\bibnamefont {Bravyi}}, \bibinfo {author} {\bibfnamefont {M.}~\bibnamefont {Motta}},\ and\ \bibinfo {author} {\bibfnamefont {G.~K.-L.}\ \bibnamefont {Chan}},\ }\bibfield  {title} {\bibinfo {title} {{Quantum Algorithms for Quantum Chemistry and Quantum Materials Science}},\ }\href@noop {} {\bibfield  {journal} {\bibinfo  {journal} {Chemical Reviews}\ }\textbf {\bibinfo {volume} {120}} (\bibinfo {year} {2020})}\BibitemShut {NoStop}%
\bibitem [{\citenamefont {Farhi}\ \emph {et~al.}(2000)\citenamefont {Farhi}, \citenamefont {Goldstone}, \citenamefont {Gutmann},\ and\ \citenamefont {Sipser}}]{farhi2000quantum}%
  \BibitemOpen
  \bibfield  {author} {\bibinfo {author} {\bibfnamefont {E.}~\bibnamefont {Farhi}}, \bibinfo {author} {\bibfnamefont {J.}~\bibnamefont {Goldstone}}, \bibinfo {author} {\bibfnamefont {S.}~\bibnamefont {Gutmann}},\ and\ \bibinfo {author} {\bibfnamefont {M.}~\bibnamefont {Sipser}},\ }\href@noop {} {\bibinfo {title} {Quantum computation by adiabatic evolution}} (\bibinfo {year} {2000}),\ \Eprint {https://arxiv.org/abs/quant-ph/0001106} {arXiv:quant-ph/0001106 [quant-ph]} \BibitemShut {NoStop}%
\bibitem [{\citenamefont {Childs}\ \emph {et~al.}(2001)\citenamefont {Childs}, \citenamefont {Farhi},\ and\ \citenamefont {Preskill}}]{Childs_2001}%
  \BibitemOpen
  \bibfield  {author} {\bibinfo {author} {\bibfnamefont {A.~M.}\ \bibnamefont {Childs}}, \bibinfo {author} {\bibfnamefont {E.}~\bibnamefont {Farhi}},\ and\ \bibinfo {author} {\bibfnamefont {J.}~\bibnamefont {Preskill}},\ }\bibfield  {title} {\bibinfo {title} {Robustness of adiabatic quantum computation},\ }\href {https://doi.org/10.1103/PhysRevA.65.012322} {\bibfield  {journal} {\bibinfo  {journal} {Phys. Rev. A}\ }\textbf {\bibinfo {volume} {65}},\ \bibinfo {pages} {012322} (\bibinfo {year} {2001})}\BibitemShut {NoStop}%
\bibitem [{\citenamefont {Farhi}\ \emph {et~al.}(2014)\citenamefont {Farhi}, \citenamefont {Goldstone},\ and\ \citenamefont {Gutmann}}]{farhiQAOA}%
  \BibitemOpen
  \bibfield  {author} {\bibinfo {author} {\bibfnamefont {E.}~\bibnamefont {Farhi}}, \bibinfo {author} {\bibfnamefont {J.}~\bibnamefont {Goldstone}},\ and\ \bibinfo {author} {\bibfnamefont {S.}~\bibnamefont {Gutmann}},\ }\bibfield  {title} {\bibinfo {title} {{A Quantum Approximate Optimization Algorithm}},\ }\href@noop {} {\bibfield  {journal} {\bibinfo  {journal} {arXiv:1411.4028}\ } (\bibinfo {year} {2014})}\BibitemShut {NoStop}%
\bibitem [{\citenamefont {Tilly}\ \emph {et~al.}(2022)\citenamefont {Tilly}, \citenamefont {Chen}, \citenamefont {Cao}, \citenamefont {Picozzi}, \citenamefont {Setia}, \citenamefont {Li}, \citenamefont {Grant}, \citenamefont {Wossnig}, \citenamefont {Rungger}, \citenamefont {Booth},\ and\ \citenamefont {Tennyson}}]{Tilly_2022}%
  \BibitemOpen
  \bibfield  {author} {\bibinfo {author} {\bibfnamefont {J.}~\bibnamefont {Tilly}}, \bibinfo {author} {\bibfnamefont {H.}~\bibnamefont {Chen}}, \bibinfo {author} {\bibfnamefont {S.}~\bibnamefont {Cao}}, \bibinfo {author} {\bibfnamefont {D.}~\bibnamefont {Picozzi}}, \bibinfo {author} {\bibfnamefont {K.}~\bibnamefont {Setia}}, \bibinfo {author} {\bibfnamefont {Y.}~\bibnamefont {Li}}, \bibinfo {author} {\bibfnamefont {E.}~\bibnamefont {Grant}}, \bibinfo {author} {\bibfnamefont {L.}~\bibnamefont {Wossnig}}, \bibinfo {author} {\bibfnamefont {I.}~\bibnamefont {Rungger}}, \bibinfo {author} {\bibfnamefont {G.~H.}\ \bibnamefont {Booth}},\ and\ \bibinfo {author} {\bibfnamefont {J.}~\bibnamefont {Tennyson}},\ }\bibfield  {title} {\bibinfo {title} {The variational quantum eigensolver: A review of methods and best practices},\ }\href {https://doi.org/10.1016/j.physrep.2022.08.003} {\bibfield  {journal} {\bibinfo  {journal} {Physics Reports}\ }\textbf {\bibinfo {volume} {986}},\ \bibinfo {pages} {1} (\bibinfo {year}
  {2022})}\BibitemShut {NoStop}%
\bibitem [{\citenamefont {Park}\ \emph {et~al.}(2016)\citenamefont {Park}, \citenamefont {Rodriguez-Briones}, \citenamefont {Feng}, \citenamefont {Rahimi}, \citenamefont {Baugh},\ and\ \citenamefont {Laflamme}}]{Park2016}%
  \BibitemOpen
  \bibfield  {author} {\bibinfo {author} {\bibfnamefont {D.~K.}\ \bibnamefont {Park}}, \bibinfo {author} {\bibfnamefont {N.~A.}\ \bibnamefont {Rodriguez-Briones}}, \bibinfo {author} {\bibfnamefont {G.}~\bibnamefont {Feng}}, \bibinfo {author} {\bibfnamefont {R.}~\bibnamefont {Rahimi}}, \bibinfo {author} {\bibfnamefont {J.}~\bibnamefont {Baugh}},\ and\ \bibinfo {author} {\bibfnamefont {R.}~\bibnamefont {Laflamme}},\ }\bibinfo {title} {{Electron Spin Resonance (ESR) Based Quantum Computing }}\ (\bibinfo  {publisher} {Springer},\ \bibinfo {year} {2016})\ Chap.\ \bibinfo {chapter} {{Heat Bath Algorithmic Cooling with Spins: Review and Prospects}}\BibitemShut {NoStop}%
\bibitem [{\citenamefont {Alhambra}\ \emph {et~al.}(2019)\citenamefont {Alhambra}, \citenamefont {Lostaglio},\ and\ \citenamefont {Perry}}]{Alhambra2019}%
  \BibitemOpen
  \bibfield  {author} {\bibinfo {author} {\bibfnamefont {A.~M.}\ \bibnamefont {Alhambra}}, \bibinfo {author} {\bibfnamefont {M.}~\bibnamefont {Lostaglio}},\ and\ \bibinfo {author} {\bibfnamefont {C.}~\bibnamefont {Perry}},\ }\bibfield  {title} {\bibinfo {title} {{Heat-Bath Algorithmic Cooling with optimal thermalization strategies}},\ }\href@noop {} {\bibfield  {journal} {\bibinfo  {journal} {Quantum}\ }\textbf {\bibinfo {volume} {3}},\ \bibinfo {pages} {188} (\bibinfo {year} {2019})}\BibitemShut {NoStop}%
\bibitem [{\citenamefont {Kaplan}\ \emph {et~al.}(2017)\citenamefont {Kaplan}, \citenamefont {Klco},\ and\ \citenamefont {Roggero}}]{Kaplan2017}%
  \BibitemOpen
  \bibfield  {author} {\bibinfo {author} {\bibfnamefont {D.~B.}\ \bibnamefont {Kaplan}}, \bibinfo {author} {\bibfnamefont {N.}~\bibnamefont {Klco}},\ and\ \bibinfo {author} {\bibfnamefont {A.}~\bibnamefont {Roggero}},\ }\href {https://arxiv.org/abs/1709.08250} {\bibinfo {title} {Ground states via spectral combing on a quantum computer}} (\bibinfo {year} {2017}),\ \Eprint {https://arxiv.org/abs/1709.08250} {arXiv:1709.08250 [quant-ph]} \BibitemShut {NoStop}%
\bibitem [{\citenamefont {Mazzola}\ \emph {et~al.}(2019)\citenamefont {Mazzola}, \citenamefont {Ollitrault}, \citenamefont {Barkoutsos},\ and\ \citenamefont {Tavernelli}}]{Mazzola2019}%
  \BibitemOpen
  \bibfield  {author} {\bibinfo {author} {\bibfnamefont {G.}~\bibnamefont {Mazzola}}, \bibinfo {author} {\bibfnamefont {P.~J.}\ \bibnamefont {Ollitrault}}, \bibinfo {author} {\bibfnamefont {P.~K.}\ \bibnamefont {Barkoutsos}},\ and\ \bibinfo {author} {\bibfnamefont {I.}~\bibnamefont {Tavernelli}},\ }\bibfield  {title} {\bibinfo {title} {Nonunitary operations for ground-state calculations in near-term quantum computers},\ }\href {https://doi.org/10.1103/PhysRevLett.123.130501} {\bibfield  {journal} {\bibinfo  {journal} {Phys. Rev. Lett.}\ }\textbf {\bibinfo {volume} {123}},\ \bibinfo {pages} {130501} (\bibinfo {year} {2019})}\BibitemShut {NoStop}%
\bibitem [{\citenamefont {Ferguson}\ \emph {et~al.}(2021)\citenamefont {Ferguson}, \citenamefont {Dellantonio}, \citenamefont {Balushi}, \citenamefont {Jansen}, \citenamefont {D\"ur},\ and\ \citenamefont {Muschik}}]{Ferguson2021}%
  \BibitemOpen
  \bibfield  {author} {\bibinfo {author} {\bibfnamefont {R.~R.}\ \bibnamefont {Ferguson}}, \bibinfo {author} {\bibfnamefont {L.}~\bibnamefont {Dellantonio}}, \bibinfo {author} {\bibfnamefont {A.~A.}\ \bibnamefont {Balushi}}, \bibinfo {author} {\bibfnamefont {K.}~\bibnamefont {Jansen}}, \bibinfo {author} {\bibfnamefont {W.}~\bibnamefont {D\"ur}},\ and\ \bibinfo {author} {\bibfnamefont {C.~A.}\ \bibnamefont {Muschik}},\ }\bibfield  {title} {\bibinfo {title} {Measurement-based variational quantum eigensolver},\ }\href {https://doi.org/10.1103/PhysRevLett.126.220501} {\bibfield  {journal} {\bibinfo  {journal} {Phys. Rev. Lett.}\ }\textbf {\bibinfo {volume} {126}},\ \bibinfo {pages} {220501} (\bibinfo {year} {2021})}\BibitemShut {NoStop}%
\bibitem [{\citenamefont {Benfenati}\ \emph {et~al.}(2021)\citenamefont {Benfenati}, \citenamefont {Mazzola}, \citenamefont {Capecci}, \citenamefont {Barkoutsos}, \citenamefont {Ollitrault}, \citenamefont {Tavernelli},\ and\ \citenamefont {Guidoni}}]{Benfenati2021}%
  \BibitemOpen
  \bibfield  {author} {\bibinfo {author} {\bibfnamefont {F.}~\bibnamefont {Benfenati}}, \bibinfo {author} {\bibfnamefont {G.}~\bibnamefont {Mazzola}}, \bibinfo {author} {\bibfnamefont {C.}~\bibnamefont {Capecci}}, \bibinfo {author} {\bibfnamefont {P.~K.}\ \bibnamefont {Barkoutsos}}, \bibinfo {author} {\bibfnamefont {P.~J.}\ \bibnamefont {Ollitrault}}, \bibinfo {author} {\bibfnamefont {I.}~\bibnamefont {Tavernelli}},\ and\ \bibinfo {author} {\bibfnamefont {L.}~\bibnamefont {Guidoni}},\ }\bibfield  {title} {\bibinfo {title} {{Improved Accuracy on Noisy Devices by Nonunitary Variational Quantum Eigensolver for Chemistry Applications}},\ }\href@noop {} {\bibfield  {journal} {\bibinfo  {journal} {J. Chem. Theory Comput.}\ }\textbf {\bibinfo {volume} {17}},\ \bibinfo {pages} {3946} (\bibinfo {year} {2021})}\BibitemShut {NoStop}%
\bibitem [{\citenamefont {Matthies}\ \emph {et~al.}(2024)\citenamefont {Matthies}, \citenamefont {Rudner}, \citenamefont {Rosch},\ and\ \citenamefont {Berg}}]{mathhies2022adibatic_demag}%
  \BibitemOpen
  \bibfield  {author} {\bibinfo {author} {\bibfnamefont {A.}~\bibnamefont {Matthies}}, \bibinfo {author} {\bibfnamefont {M.}~\bibnamefont {Rudner}}, \bibinfo {author} {\bibfnamefont {A.}~\bibnamefont {Rosch}},\ and\ \bibinfo {author} {\bibfnamefont {E.}~\bibnamefont {Berg}},\ }\bibfield  {title} {\bibinfo {title} {Programmable adiabatic demagnetization for systems with trivial and topological excitations},\ }\href {https://quantum-journal.org/papers/q-2024-10-23-1505/} {\bibfield  {journal} {\bibinfo  {journal} {Quantum}\ }\textbf {\bibinfo {volume} {8}} (\bibinfo {year} {2024})}\BibitemShut {NoStop}%
\bibitem [{\citenamefont {Cubitt}(2023)}]{Cubitt2023}%
  \BibitemOpen
  \bibfield  {author} {\bibinfo {author} {\bibfnamefont {T.~S.}\ \bibnamefont {Cubitt}},\ }\bibfield  {title} {\bibinfo {title} {{Dissipative ground state preparation and the Dissipative Quantum Eigensolver}},\ }\href@noop {} {\bibfield  {journal} {\bibinfo  {journal} {arXiv:2303.11962}\ } (\bibinfo {year} {2023})}\BibitemShut {NoStop}%
\bibitem [{\citenamefont {Piroli}\ \emph {et~al.}(2024)\citenamefont {Piroli}, \citenamefont {Styliaris},\ and\ \citenamefont {Cirac}}]{Piroli2024}%
  \BibitemOpen
  \bibfield  {author} {\bibinfo {author} {\bibfnamefont {L.}~\bibnamefont {Piroli}}, \bibinfo {author} {\bibfnamefont {G.}~\bibnamefont {Styliaris}},\ and\ \bibinfo {author} {\bibfnamefont {J.~I.}\ \bibnamefont {Cirac}},\ }\bibfield  {title} {\bibinfo {title} {Approximating many-body quantum states with quantum circuits and measurements},\ }\href {https://doi.org/10.1103/PhysRevLett.133.230401} {\bibfield  {journal} {\bibinfo  {journal} {Phys. Rev. Lett.}\ }\textbf {\bibinfo {volume} {133}},\ \bibinfo {pages} {230401} (\bibinfo {year} {2024})}\BibitemShut {NoStop}%
\bibitem [{\citenamefont {Cobos}\ \emph {et~al.}(2024)\citenamefont {Cobos}, \citenamefont {Locher}, \citenamefont {Bermudez}, \citenamefont {M\"uller},\ and\ \citenamefont {Rico}}]{Cobos2024}%
  \BibitemOpen
  \bibfield  {author} {\bibinfo {author} {\bibfnamefont {J.}~\bibnamefont {Cobos}}, \bibinfo {author} {\bibfnamefont {D.~F.}\ \bibnamefont {Locher}}, \bibinfo {author} {\bibfnamefont {A.}~\bibnamefont {Bermudez}}, \bibinfo {author} {\bibfnamefont {M.}~\bibnamefont {M\"uller}},\ and\ \bibinfo {author} {\bibfnamefont {E.}~\bibnamefont {Rico}},\ }\bibfield  {title} {\bibinfo {title} {Noise-aware variational eigensolvers: A dissipative route for lattice gauge theories},\ }\href {https://doi.org/10.1103/PRXQuantum.5.030340} {\bibfield  {journal} {\bibinfo  {journal} {PRX Quantum}\ }\textbf {\bibinfo {volume} {5}},\ \bibinfo {pages} {030340} (\bibinfo {year} {2024})}\BibitemShut {NoStop}%
\bibitem [{\citenamefont {Molpeceres}\ \emph {et~al.}(2025)\citenamefont {Molpeceres}, \citenamefont {Lu}, \citenamefont {Cirac},\ and\ \citenamefont {Kraus}}]{Molpeceres2025}%
  \BibitemOpen
  \bibfield  {author} {\bibinfo {author} {\bibfnamefont {D.}~\bibnamefont {Molpeceres}}, \bibinfo {author} {\bibfnamefont {S.}~\bibnamefont {Lu}}, \bibinfo {author} {\bibfnamefont {J.~I.}\ \bibnamefont {Cirac}},\ and\ \bibinfo {author} {\bibfnamefont {B.}~\bibnamefont {Kraus}},\ }\bibfield  {title} {\bibinfo {title} {Quantum algorithms for cooling: A simple case study},\ }\href {https://doi.org/10.1103/4hx7-xnhw} {\bibfield  {journal} {\bibinfo  {journal} {Phys. Rev. Res.}\ }\textbf {\bibinfo {volume} {7}},\ \bibinfo {pages} {033162} (\bibinfo {year} {2025})}\BibitemShut {NoStop}%
\bibitem [{\citenamefont {Kishony}\ \emph {et~al.}(2025{\natexlab{a}})\citenamefont {Kishony}, \citenamefont {Rudner}, \citenamefont {Rosch},\ and\ \citenamefont {Berg}}]{kishony2025gauged}%
  \BibitemOpen
  \bibfield  {author} {\bibinfo {author} {\bibfnamefont {G.}~\bibnamefont {Kishony}}, \bibinfo {author} {\bibfnamefont {M.~S.}\ \bibnamefont {Rudner}}, \bibinfo {author} {\bibfnamefont {A.}~\bibnamefont {Rosch}},\ and\ \bibinfo {author} {\bibfnamefont {E.}~\bibnamefont {Berg}},\ }\bibfield  {title} {\bibinfo {title} {Gauged cooling of topological excitations and emergent fermions on quantum simulators},\ }\href {https://doi.org/10.1103/PhysRevLett.134.086503} {\bibfield  {journal} {\bibinfo  {journal} {Phys. Rev. Lett.}\ }\textbf {\bibinfo {volume} {134}},\ \bibinfo {pages} {086503} (\bibinfo {year} {2025}{\natexlab{a}})}\BibitemShut {NoStop}%
\bibitem [{\citenamefont {Kishony}\ \emph {et~al.}(2025{\natexlab{b}})\citenamefont {Kishony}, \citenamefont {Rudner},\ and\ \citenamefont {Berg}}]{kishony2025_chiral}%
  \BibitemOpen
  \bibfield  {author} {\bibinfo {author} {\bibfnamefont {G.}~\bibnamefont {Kishony}}, \bibinfo {author} {\bibfnamefont {M.~S.}\ \bibnamefont {Rudner}},\ and\ \bibinfo {author} {\bibfnamefont {E.}~\bibnamefont {Berg}},\ }\bibfield  {title} {\bibinfo {title} {Efficiently preparing chiral states via fermionic cooling on bosonic quantum hardware},\ }\href {https://doi.org/https://doi.org/10.1038/s42005-025-02002-7} {\bibfield  {journal} {\bibinfo  {journal} {Comms. Phys.}\ }\textbf {\bibinfo {volume} {8}},\ \bibinfo {pages} {73} (\bibinfo {year} {2025}{\natexlab{b}})}\BibitemShut {NoStop}%
\bibitem [{\citenamefont {Mi}\ \emph {et~al.}(2024)\citenamefont {Mi}, \citenamefont {Michailidis}, \citenamefont {Shabani}, \citenamefont {Miao}, \citenamefont {Klimov}, \citenamefont {Lloyd}, \citenamefont {Rosenberg} \emph {et~al.}}]{mi2023stable}%
  \BibitemOpen
  \bibfield  {author} {\bibinfo {author} {\bibfnamefont {X.}~\bibnamefont {Mi}}, \bibinfo {author} {\bibfnamefont {A.~A.}\ \bibnamefont {Michailidis}}, \bibinfo {author} {\bibfnamefont {S.}~\bibnamefont {Shabani}}, \bibinfo {author} {\bibfnamefont {K.~C.}\ \bibnamefont {Miao}}, \bibinfo {author} {\bibfnamefont {P.~V.}\ \bibnamefont {Klimov}}, \bibinfo {author} {\bibfnamefont {J.}~\bibnamefont {Lloyd}}, \bibinfo {author} {\bibfnamefont {E.}~\bibnamefont {Rosenberg}}, \emph {et~al.},\ }\bibfield  {title} {\bibinfo {title} {Stable quantum-correlated many body states via engineered dissipation},\ }\href@noop {} {\bibfield  {journal} {\bibinfo  {journal} {Science}\ }\textbf {\bibinfo {volume} {383}},\ \bibinfo {pages} {1332} (\bibinfo {year} {2024})}\BibitemShut {NoStop}%
\bibitem [{\citenamefont {Lloyd}\ \emph {et~al.}(2025)\citenamefont {Lloyd}, \citenamefont {Michailidis}, \citenamefont {Mi}, \citenamefont {Smelyanskiy},\ and\ \citenamefont {Abanin}}]{Lloyd2025}%
  \BibitemOpen
  \bibfield  {author} {\bibinfo {author} {\bibfnamefont {J.}~\bibnamefont {Lloyd}}, \bibinfo {author} {\bibfnamefont {A.~A.}\ \bibnamefont {Michailidis}}, \bibinfo {author} {\bibfnamefont {X.}~\bibnamefont {Mi}}, \bibinfo {author} {\bibfnamefont {V.}~\bibnamefont {Smelyanskiy}},\ and\ \bibinfo {author} {\bibfnamefont {D.~A.}\ \bibnamefont {Abanin}},\ }\bibfield  {title} {\bibinfo {title} {Quasiparticle cooling algorithms for quantum many-body state preparation},\ }\href {https://doi.org/10.1103/PRXQuantum.6.010361} {\bibfield  {journal} {\bibinfo  {journal} {PRX Quantum}\ }\textbf {\bibinfo {volume} {6}},\ \bibinfo {pages} {010361} (\bibinfo {year} {2025})}\BibitemShut {NoStop}%
\bibitem [{\citenamefont {Ding}\ \emph {et~al.}(2024)\citenamefont {Ding}, \citenamefont {Chen},\ and\ \citenamefont {Lin}}]{Ding2024}%
  \BibitemOpen
  \bibfield  {author} {\bibinfo {author} {\bibfnamefont {Z.}~\bibnamefont {Ding}}, \bibinfo {author} {\bibfnamefont {C.-F.}\ \bibnamefont {Chen}},\ and\ \bibinfo {author} {\bibfnamefont {L.}~\bibnamefont {Lin}},\ }\bibfield  {title} {\bibinfo {title} {{Single-ancilla ground state preparation via Lindbladians}},\ }\href {https://doi.org/10.1103/PhysRevResearch.6.033147} {\bibfield  {journal} {\bibinfo  {journal} {Phys. Rev. Res.}\ }\textbf {\bibinfo {volume} {6}},\ \bibinfo {pages} {033147} (\bibinfo {year} {2024})}\BibitemShut {NoStop}%
\bibitem [{\citenamefont {Ding}\ \emph {et~al.}(2025{\natexlab{a}})\citenamefont {Ding}, \citenamefont {Zhan}, \citenamefont {Preskill},\ and\ \citenamefont {Lin}}]{DingEndtoEnd}%
  \BibitemOpen
  \bibfield  {author} {\bibinfo {author} {\bibfnamefont {Z.}~\bibnamefont {Ding}}, \bibinfo {author} {\bibfnamefont {Y.}~\bibnamefont {Zhan}}, \bibinfo {author} {\bibfnamefont {J.}~\bibnamefont {Preskill}},\ and\ \bibinfo {author} {\bibfnamefont {L.}~\bibnamefont {Lin}},\ }\bibfield  {title} {\bibinfo {title} {{End-to-End Efficient Quantum Thermal and Ground State Preparation Made Simple}},\ }\href@noop {} {\bibfield  {journal} {\bibinfo  {journal} {arXiv:2508.05703}\ } (\bibinfo {year} {2025}{\natexlab{a}})}\BibitemShut {NoStop}%
\bibitem [{\citenamefont {Zhan}\ \emph {et~al.}(2025)\citenamefont {Zhan}, \citenamefont {Ding}, \citenamefont {Huhn}, \citenamefont {Gray}, \citenamefont {Preskill}, \citenamefont {Chan},\ and\ \citenamefont {Lin}}]{Zhan2025}%
  \BibitemOpen
  \bibfield  {author} {\bibinfo {author} {\bibfnamefont {Y.}~\bibnamefont {Zhan}}, \bibinfo {author} {\bibfnamefont {Z.}~\bibnamefont {Ding}}, \bibinfo {author} {\bibfnamefont {J.}~\bibnamefont {Huhn}}, \bibinfo {author} {\bibfnamefont {J.}~\bibnamefont {Gray}}, \bibinfo {author} {\bibfnamefont {J.}~\bibnamefont {Preskill}}, \bibinfo {author} {\bibfnamefont {G.~K.-L.}\ \bibnamefont {Chan}},\ and\ \bibinfo {author} {\bibfnamefont {L.}~\bibnamefont {Lin}},\ }\bibfield  {title} {\bibinfo {title} {{Rapid quantum ground state preparation via dissipative dynamics}},\ }\href@noop {} {\bibfield  {journal} {\bibinfo  {journal} {arXiv:2503.15827}\ } (\bibinfo {year} {2025})}\BibitemShut {NoStop}%
\bibitem [{\citenamefont {Marti}\ \emph {et~al.}(2025)\citenamefont {Marti}, \citenamefont {Mansuroglu},\ and\ \citenamefont {Hartmann}}]{Marti2025}%
  \BibitemOpen
  \bibfield  {author} {\bibinfo {author} {\bibfnamefont {L.}~\bibnamefont {Marti}}, \bibinfo {author} {\bibfnamefont {R.}~\bibnamefont {Mansuroglu}},\ and\ \bibinfo {author} {\bibfnamefont {M.~J.}\ \bibnamefont {Hartmann}},\ }\bibfield  {title} {\bibinfo {title} {{Efficient Quantum Cooling Algorithm for Fermionic Systems}},\ }\href@noop {} {\bibfield  {journal} {\bibinfo  {journal} {Quantum}\ }\textbf {\bibinfo {volume} {9}},\ \bibinfo {pages} {1635} (\bibinfo {year} {2025})}\BibitemShut {NoStop}%
\bibitem [{\citenamefont {Langbehn}\ \emph {et~al.}(2025)\citenamefont {Langbehn}, \citenamefont {Mouloudakis}, \citenamefont {King}, \citenamefont {Menu}, \citenamefont {Gornyi}, \citenamefont {Morigi}, \citenamefont {Gefen},\ and\ \citenamefont {Koch}}]{langbehn2025}%
  \BibitemOpen
  \bibfield  {author} {\bibinfo {author} {\bibfnamefont {J.}~\bibnamefont {Langbehn}}, \bibinfo {author} {\bibfnamefont {G.}~\bibnamefont {Mouloudakis}}, \bibinfo {author} {\bibfnamefont {E.}~\bibnamefont {King}}, \bibinfo {author} {\bibfnamefont {R.}~\bibnamefont {Menu}}, \bibinfo {author} {\bibfnamefont {I.}~\bibnamefont {Gornyi}}, \bibinfo {author} {\bibfnamefont {G.}~\bibnamefont {Morigi}}, \bibinfo {author} {\bibfnamefont {Y.}~\bibnamefont {Gefen}},\ and\ \bibinfo {author} {\bibfnamefont {C.~P.}\ \bibnamefont {Koch}},\ }\href {https://arxiv.org/abs/2506.11964} {\bibinfo {title} {Universal cooling of quantum systems via randomized measurements}} (\bibinfo {year} {2025}),\ \Eprint {https://arxiv.org/abs/2506.11964} {arXiv:2506.11964 [quant-ph]} \BibitemShut {NoStop}%
\bibitem [{\citenamefont {Wiersema}\ \emph {et~al.}(2020)\citenamefont {Wiersema}, \citenamefont {Zhou}, \citenamefont {de~Sereville}, \citenamefont {Carrasquilla}, \citenamefont {Kim},\ and\ \citenamefont {Yuen}}]{Wiersema2020}%
  \BibitemOpen
  \bibfield  {author} {\bibinfo {author} {\bibfnamefont {R.}~\bibnamefont {Wiersema}}, \bibinfo {author} {\bibfnamefont {C.}~\bibnamefont {Zhou}}, \bibinfo {author} {\bibfnamefont {Y.}~\bibnamefont {de~Sereville}}, \bibinfo {author} {\bibfnamefont {J.~F.}\ \bibnamefont {Carrasquilla}}, \bibinfo {author} {\bibfnamefont {Y.~B.}\ \bibnamefont {Kim}},\ and\ \bibinfo {author} {\bibfnamefont {H.}~\bibnamefont {Yuen}},\ }\bibfield  {title} {\bibinfo {title} {{Exploring Entanglement and Optimization within the Hamiltonian Variational Ansatz}},\ }\href {https://doi.org/10.1103/PRXQuantum.1.020319} {\bibfield  {journal} {\bibinfo  {journal} {PRX Quantum}\ }\textbf {\bibinfo {volume} {1}},\ \bibinfo {pages} {020319} (\bibinfo {year} {2020})}\BibitemShut {NoStop}%
\bibitem [{IBM(2025{\natexlab{a}})}]{IBM_Quantum}%
  \BibitemOpen
  \href@noop {} {\bibinfo {title} {For information on the quantum processor, see {IBM Quantum Platform, https://quantum.cloud.ibm.com/}}} (\bibinfo {year} {2025}{\natexlab{a}})\BibitemShut {NoStop}%
\bibitem [{\citenamefont {Shin}\ \emph {et~al.}(2025)\citenamefont {Shin}, \citenamefont {Kim}, \citenamefont {Yeo}, \citenamefont {Jeong}, \citenamefont {Jhe},\ and\ \citenamefont {Kim}}]{Shin2025}%
  \BibitemOpen
  \bibfield  {author} {\bibinfo {author} {\bibfnamefont {S.}~\bibnamefont {Shin}}, \bibinfo {author} {\bibfnamefont {H.~E.}\ \bibnamefont {Kim}}, \bibinfo {author} {\bibfnamefont {H.}~\bibnamefont {Yeo}}, \bibinfo {author} {\bibfnamefont {K.}~\bibnamefont {Jeong}}, \bibinfo {author} {\bibfnamefont {W.}~\bibnamefont {Jhe}},\ and\ \bibinfo {author} {\bibfnamefont {J.}~\bibnamefont {Kim}},\ }\bibfield  {title} {\bibinfo {title} {{Designing Minimalistic Variational Quantum Ansatz Inspired by Algorithmic Cooling}},\ }\href@noop {} {\bibfield  {journal} {\bibinfo  {journal} {arXiv:2501.16776}\ } (\bibinfo {year} {2025})}\BibitemShut {NoStop}%
\bibitem [{Note1()}]{Note1}%
  \BibitemOpen
  \bibinfo {note} {Note that the experiment of Ref.~\cite {mi2023stable} employed variational optimization with respect to the values of a constant (layer-independent) applied bath field and a constant system-bath coupling strength. The Vari-Cool protocol is a general purpose variational scheme in which the unitary block may or may not resemble Trotterized evolution, and where all parameters may vary independently on every layer.}\BibitemShut {Stop}%
\bibitem [{Note2()}]{Note2}%
  \BibitemOpen
  \bibinfo {note} {Ideally, $n_{\protect \rm bath} = N$ would be more effective for cooling the system qubits. However, the additional overhead that would be needed to introduce bath couplings to every system site with the restricted connectivity shown in Fig.~\ref {fig:QPU} would far outweigh the benefits.}\BibitemShut {Stop}%
\bibitem [{\citenamefont {Cerezo}\ \emph {et~al.}(2021)\citenamefont {Cerezo}, \citenamefont {Arrasmith}, \citenamefont {Babbush}, \citenamefont {Bejamin}, \citenamefont {Endo}, \citenamefont {Fujii}, \citenamefont {McClean}, \citenamefont {Mitarai}, \citenamefont {Yuan}, \citenamefont {Cincio},\ and\ \citenamefont {Coles}}]{Cerezo2021}%
  \BibitemOpen
  \bibfield  {author} {\bibinfo {author} {\bibfnamefont {M.}~\bibnamefont {Cerezo}}, \bibinfo {author} {\bibfnamefont {A.}~\bibnamefont {Arrasmith}}, \bibinfo {author} {\bibfnamefont {R.}~\bibnamefont {Babbush}}, \bibinfo {author} {\bibfnamefont {S.~C.}\ \bibnamefont {Bejamin}}, \bibinfo {author} {\bibfnamefont {S.}~\bibnamefont {Endo}}, \bibinfo {author} {\bibfnamefont {K.}~\bibnamefont {Fujii}}, \bibinfo {author} {\bibfnamefont {J.~R.}\ \bibnamefont {McClean}}, \bibinfo {author} {\bibfnamefont {K.}~\bibnamefont {Mitarai}}, \bibinfo {author} {\bibfnamefont {X.}~\bibnamefont {Yuan}}, \bibinfo {author} {\bibfnamefont {L.}~\bibnamefont {Cincio}},\ and\ \bibinfo {author} {\bibfnamefont {P.~J.}\ \bibnamefont {Coles}},\ }\bibfield  {title} {\bibinfo {title} {{Variational quantum algorithms}},\ }\href@noop {} {\bibfield  {journal} {\bibinfo  {journal} {Nature Reviews Physics}\ ,\ \bibinfo {pages} {625}} (\bibinfo {year} {2021})}\BibitemShut {NoStop}%
\bibitem [{Note3()}]{Note3}%
  \BibitemOpen
  \bibinfo {note} {The Nelder-Mead method is a gradient-free approach for minimizing or maximizing multidimensional cost functions. See \protect \href {https://docs.scipy.org/doc/scipy/reference/optimize.minimize-neldermead.html}{SciPy documentation} and reference therein for further information.}\BibitemShut {Stop}%
\bibitem [{Note4()}]{Note4}%
  \BibitemOpen
  \bibinfo {note} {Here it is important to also enforce a monotonic decrease of energy with cycle number around $T_{\protect \rm train}$ to avoid unsteady behavior, as we impose throughout the optimization.}\BibitemShut {Stop}%
\bibitem [{\citenamefont {Javadi-Abhari}\ \emph {et~al.}(2024)\citenamefont {Javadi-Abhari}, \citenamefont {Treinish}, \citenamefont {Krsulich}, \citenamefont {Wood}, \citenamefont {Lishman}, \citenamefont {Gacon}, \citenamefont {Martiel}, \citenamefont {Nation}, \citenamefont {Bishop}, \citenamefont {Cross}, \citenamefont {Johnson},\ and\ \citenamefont {Gambetta}}]{javadiabhari2024quantumcomputingqiskit}%
  \BibitemOpen
  \bibfield  {author} {\bibinfo {author} {\bibfnamefont {A.}~\bibnamefont {Javadi-Abhari}}, \bibinfo {author} {\bibfnamefont {M.}~\bibnamefont {Treinish}}, \bibinfo {author} {\bibfnamefont {K.}~\bibnamefont {Krsulich}}, \bibinfo {author} {\bibfnamefont {C.~J.}\ \bibnamefont {Wood}}, \bibinfo {author} {\bibfnamefont {J.}~\bibnamefont {Lishman}}, \bibinfo {author} {\bibfnamefont {J.}~\bibnamefont {Gacon}}, \bibinfo {author} {\bibfnamefont {S.}~\bibnamefont {Martiel}}, \bibinfo {author} {\bibfnamefont {P.~D.}\ \bibnamefont {Nation}}, \bibinfo {author} {\bibfnamefont {L.~S.}\ \bibnamefont {Bishop}}, \bibinfo {author} {\bibfnamefont {A.~W.}\ \bibnamefont {Cross}}, \bibinfo {author} {\bibfnamefont {B.~R.}\ \bibnamefont {Johnson}},\ and\ \bibinfo {author} {\bibfnamefont {J.~M.}\ \bibnamefont {Gambetta}},\ }\href {https://arxiv.org/abs/2405.08810} {\bibinfo {title} {{Quantum computing with Qiskit}}} (\bibinfo {year} {2024}),\ \Eprint {https://arxiv.org/abs/2405.08810} {arXiv:2405.08810 [quant-ph]} \BibitemShut
  {NoStop}%
\bibitem [{Note5()}]{Note5}%
  \BibitemOpen
  \bibinfo {note} {Each $\protect \hat {R}_{yy}(\theta )$ gate in the protocol is transpiled to a native $\protect \hat {R}_{zz}(\theta )$ gate and single qubit gates.}\BibitemShut {Stop}%
\bibitem [{Note6()}]{Note6}%
  \BibitemOpen
  \bibinfo {note} {We have not systematically studied the effects of cross-talk during resets, but did find overall that the two qubits per bath site approach described in Fig.~\ref {fig:QPU} gave better results than a naive direct implementation with one qubit per bath site.}\BibitemShut {Stop}%
\bibitem [{Note7()}]{Note7}%
  \BibitemOpen
  \bibinfo {note} {{\protect \tt SWAP} gates are not native to the {\protect \tt ibm\protect \_kingston} processor. Instead, a {\protect \tt SWAP} operation can be transpiled to a sequence of three {\protect \tt CX} gates acting in alternating directions between the two qubits. In our protocol, however, one of the qubits involved in each {\protect \tt SWAP} operation always begins in the $|0\rangle $ state, rendering the first of the three {\protect \tt CX} gates redundant. Thus only two {\protect \tt CX} gates are required. Finally, each {\protect \tt CX} gate can be implemented using one native {\protect \tt CZ} gate and single qubit gates.}\BibitemShut {Stop}%
\bibitem [{\citenamefont {Rall}\ \emph {et~al.}(2023)\citenamefont {Rall}, \citenamefont {Wang},\ and\ \citenamefont {Wocjan}}]{Rall2023}%
  \BibitemOpen
  \bibfield  {author} {\bibinfo {author} {\bibfnamefont {P.}~\bibnamefont {Rall}}, \bibinfo {author} {\bibfnamefont {C.}~\bibnamefont {Wang}},\ and\ \bibinfo {author} {\bibfnamefont {P.}~\bibnamefont {Wocjan}},\ }\bibfield  {title} {\bibinfo {title} {{Thermal State Preparation via Rounding Promises}},\ }\href@noop {} {\bibfield  {journal} {\bibinfo  {journal} {Quantum}\ }\textbf {\bibinfo {volume} {7}},\ \bibinfo {pages} {1132} (\bibinfo {year} {2023})}\BibitemShut {NoStop}%
\bibitem [{\citenamefont {Chen}\ \emph {et~al.}(2023{\natexlab{a}})\citenamefont {Chen}, \citenamefont {Kastoryano}, \citenamefont {Brandão},\ and\ \citenamefont {Gilyen}}]{Chen2023}%
  \BibitemOpen
  \bibfield  {author} {\bibinfo {author} {\bibfnamefont {C.-F.}\ \bibnamefont {Chen}}, \bibinfo {author} {\bibfnamefont {M.~J.}\ \bibnamefont {Kastoryano}}, \bibinfo {author} {\bibfnamefont {F.~G.}\ \bibnamefont {Brandão}},\ and\ \bibinfo {author} {\bibfnamefont {A.}~\bibnamefont {Gilyen}},\ }\bibfield  {title} {\bibinfo {title} {{Quantum Thermal State Preparation}},\ }\href@noop {} {\bibfield  {journal} {\bibinfo  {journal} {arXiv:2023.18224}\ } (\bibinfo {year} {2023}{\natexlab{a}})}\BibitemShut {NoStop}%
\bibitem [{\citenamefont {Chen}\ \emph {et~al.}(2023{\natexlab{b}})\citenamefont {Chen}, \citenamefont {Kastoryano},\ and\ \citenamefont {Gilyen}}]{Chen2023EfficientGibbs}%
  \BibitemOpen
  \bibfield  {author} {\bibinfo {author} {\bibfnamefont {C.-F.}\ \bibnamefont {Chen}}, \bibinfo {author} {\bibfnamefont {M.~J.}\ \bibnamefont {Kastoryano}},\ and\ \bibinfo {author} {\bibfnamefont {A.}~\bibnamefont {Gilyen}},\ }\bibfield  {title} {\bibinfo {title} {{An efficient and exact noncommutative quantum Gibbs sampler}},\ }\href@noop {} {\bibfield  {journal} {\bibinfo  {journal} {arXiv:2311.09207}\ } (\bibinfo {year} {2023}{\natexlab{b}})}\BibitemShut {NoStop}%
\bibitem [{\citenamefont {Ding}\ \emph {et~al.}(2025{\natexlab{b}})\citenamefont {Ding}, \citenamefont {Li},\ and\ \citenamefont {Lin}}]{Ding2025}%
  \BibitemOpen
  \bibfield  {author} {\bibinfo {author} {\bibfnamefont {Z.}~\bibnamefont {Ding}}, \bibinfo {author} {\bibfnamefont {B.}~\bibnamefont {Li}},\ and\ \bibinfo {author} {\bibfnamefont {L.}~\bibnamefont {Lin}},\ }\bibfield  {title} {\bibinfo {title} {{Efficient quantum Gibbs samplers with Kubo--Martin--Schwinger detailed balance condition}},\ }\href@noop {} {\bibfield  {journal} {\bibinfo  {journal} {Commun. Math. Phys.}\ }\textbf {\bibinfo {volume} {406}},\ \bibinfo {pages} {67} (\bibinfo {year} {2025}{\natexlab{b}})}\BibitemShut {NoStop}%
\bibitem [{\citenamefont {Guo}\ \emph {et~al.}(2025)\citenamefont {Guo}, \citenamefont {Hart}, \citenamefont {Chen}, \citenamefont {Friedman},\ and\ \citenamefont {Lucas}}]{Guo2025}%
  \BibitemOpen
  \bibfield  {author} {\bibinfo {author} {\bibfnamefont {J.}~\bibnamefont {Guo}}, \bibinfo {author} {\bibfnamefont {O.}~\bibnamefont {Hart}}, \bibinfo {author} {\bibfnamefont {C.-F.}\ \bibnamefont {Chen}}, \bibinfo {author} {\bibfnamefont {A.~J.}\ \bibnamefont {Friedman}},\ and\ \bibinfo {author} {\bibfnamefont {A.}~\bibnamefont {Lucas}},\ }\bibfield  {title} {\bibinfo {title} {{Designing open quantum systems with known steady states: Davies generators and beyond}},\ }\href@noop {} {\bibfield  {journal} {\bibinfo  {journal} {Quantum}\ }\textbf {\bibinfo {volume} {9}},\ \bibinfo {pages} {1612} (\bibinfo {year} {2025})}\BibitemShut {NoStop}%
\bibitem [{\citenamefont {Hahn}\ \emph {et~al.}(2025)\citenamefont {Hahn}, \citenamefont {Parameswaran},\ and\ \citenamefont {Placke}}]{Hahn2025}%
  \BibitemOpen
  \bibfield  {author} {\bibinfo {author} {\bibfnamefont {D.}~\bibnamefont {Hahn}}, \bibinfo {author} {\bibfnamefont {S.}~\bibnamefont {Parameswaran}},\ and\ \bibinfo {author} {\bibfnamefont {B.}~\bibnamefont {Placke}},\ }\bibfield  {title} {\bibinfo {title} {{Provably efficient quantum thermal state preparation via local driving}},\ }\href@noop {} {\bibfield  {journal} {\bibinfo  {journal} {arXiv:2505.22816}\ } (\bibinfo {year} {2025})}\BibitemShut {NoStop}%
\bibitem [{\citenamefont {Lloyd}\ and\ \citenamefont {Abanin}(2025)}]{Lloyd2025Thermal}%
  \BibitemOpen
  \bibfield  {author} {\bibinfo {author} {\bibfnamefont {J.}~\bibnamefont {Lloyd}}\ and\ \bibinfo {author} {\bibfnamefont {D.~A.}\ \bibnamefont {Abanin}},\ }\bibfield  {title} {\bibinfo {title} {{Quantum thermal state preparation for near-term quantum processors}},\ }\href@noop {} {\bibfield  {journal} {\bibinfo  {journal} {arXiv:2506.21318}\ } (\bibinfo {year} {2025})}\BibitemShut {NoStop}%
\bibitem [{Note8()}]{Note8}%
  \BibitemOpen
  \bibinfo {note} {We note that initial guesses populated with many $0$ values do not give much cooling, and thus typically do not provide a good starting point for optimization (cf.~Ref.~\protect \rev@citealp {Cerezo2021}).}\BibitemShut {Stop}%
\bibitem [{IBM(2025{\natexlab{b}})}]{IBM_Quantum_ErrorDefs}%
  \BibitemOpen
  \href@noop {} {\bibinfo {title} {{For definitions of how gate and readout errors are defined, see QPU Information on IBM Quantum Platform, https://quantum.cloud.ibm.com/docs/en/guides/qpu-information}}} (\bibinfo {year} {2025}{\natexlab{b}})\BibitemShut {NoStop}%
\end{thebibliography}%

\end{document}